\documentclass[11pt,a4paper]{article}
\pdfoutput=1

\usepackage{jheppub}
\usepackage{latexsym}
\usepackage{multirow}
\usepackage{color}
\usepackage[usenames,dvipsnames,table]{xcolor}
\usepackage{graphicx}
\usepackage{epsfig}  
\usepackage{dcolumn}
\usepackage{bm}
\usepackage{dcolumn}
\usepackage{textcomp}
\usepackage{float}
\usepackage{hypcap}
\usepackage[]{hyperref}
\usepackage{makecell}
\usepackage{multirow}
\usepackage{color}
\usepackage{pifont}
\usepackage{subfigure}
\usepackage{amssymb}
 \usepackage{caption}
  \newcommand{\capdef}{}
  \newcommand{\mycaption}[2][\capdef]{\renewcommand{\capdef}{#2}
       \caption[#1]{{\footnotesize #2}}}
  \newcommand{\be}{\begin{equation}}
   \newcommand{\ee}{\end{equation}}

\hypersetup{
  bookmarks=true,         
  unicode=false,          
  pdftoolbar=true,        
 pdfmenubar=true,        
 pdffitwindow=true,     
 pdfstartview={FitH},    
 pdfsubject={Neutrino Oscillations Phenomenology},   
 pdfnewwindow=true,      
 pdfcreator={RevTeX},
 colorlinks=true,       
 linkcolor=red,          
 citecolor=blue,        
 filecolor=black,      
 urlcolor=blue,           
  }

\title{Can INO be Sensitive to Flavor-Dependent Long-Range Forces?}

\author[a,b]{Amina Khatun,}
\author[c]{Tarak Thakore,}
\author[a,b]{Sanjib Kumar Agarwalla}

\affiliation[a]{Institute of Physics, Sachivalaya Marg, Sainik School Post, Bhubaneswar 751005, India}
\affiliation[b]{Homi Bhabha National Institute, Training School Complex, Anushakti Nagar, Mumbai 400085, India}
\affiliation[c]{Louisiana State University, Baton Rouge, Louisiana, 70803 U.S.A.}

\emailAdd{amina@iopb.res.in}
\emailAdd{thakore@phys.lsu.edu}
\emailAdd{sanjib@iopb.res.in}

 \abstract{Flavor-dependent long-range leptonic forces mediated by the ultra-light and neutral 
 bosons associated with gauged $L_e-L_\mu$ or $L_e-L_\tau$ symmetry constitute a minimal 
 extension of the Standard Model. In presence of these new anomaly free abelian symmetries, the SM 
 remains invariant and renormalizable, and can lead to interesting phenomenological consequences. 
 For an example, the electrons inside the Sun can generate a flavor-dependent long-range potential 
 at the Earth surface, which can enhance $\nu_\mu$ and $\bar\nu_\mu$ survival probabilities 
 over a wide range of energies and baselines in atmospheric neutrino experiments.  In this paper,  
 we explore in detail the possible impacts of these long-range flavor-diagonal neutral current 
 interactions due to $L_e-L_\mu$ and  $L_e-L_\tau$ symmetries (one at-a-time) in the context of  
 proposed 50 kt magnetized ICAL detector at INO. Combining the information on muon momentum and 
 hadron energy on an event-by-event basis, ICAL can place  stringent constraints on the 
 effective gauge coupling $\alpha_{e\mu/e\tau}<1.2\times 10^{-53}$ ($1.75\times 
 10^{-53}$) at 90$\%$ (3$\sigma$)  C.L. with 500 kt$\cdot$yr exposure. 
 The 90$\%$  C.L. limit on $\alpha_{e\mu}$ ($\alpha_{e\tau}$) from ICAL is $\sim 46$ (53) times better 
 than the existing bound from the Super-Kamiokande experiment.
 }

\arxivnumber{1801.aaaaa}

\begin{document}
\preprint{IP/BBSR/2017-14}
\maketitle
\flushbottom

\section{Introduction and Motivation}
The confirmation of neutrino flavor oscillation via several pioneering experiments
over the last two decades is a landmark achievement in the intensity frontier of the 
high energy particle physics\,\cite{Olive:2016xmw}. All the neutrino oscillation data 
available so far can be accommodated in the standard three-flavor oscillation picture 
of neutrinos\,\cite{Esteban:2016qun,deSalas:2017kay,Capozzi:2017ipn}. The 3$\nu$ 
mixing framework contains six fundamental parameters: a) three mixing angles ($\theta_{12}$, $\theta_{13}$,
$\theta_{23}$), b) one Dirac CP phase ($\delta_{\rm CP}$), and c) two independent mass-squared
differences\footnote{In the solar sector, we have $\Delta m^2_{21}\equiv m^2_{2}-m^2_{1}$ and 
in the atmospheric sector, we deal with $\Delta m^2_{32} \equiv m^2_{3}-m^2_{2}$, where 
$m_{3}$ corresponds to the neutrino mass eigenstate with the smallest electron component.} 
($\Delta m^2_{21}$ and $\Delta m^2_{32}$).

Let us briefly discuss about the present status of these oscillation parameters.  
According to the latest global fit of world neutrino data 
available till November 2017\,\cite{Esteban:2016qun,NuFIT}, the best fit values of the solar 
parameters $\sin^2\theta_{12}$ and $\Delta m^2_{21}$ are 0.307 and 7.4$\times 10^{-5}$ eV$^2$ 
respectively.  
The relative 1$\sigma$ precision\footnote{Here, 1$\sigma$ precision is defined as 1/6 of 
the $\pm 3 \sigma$ range. } on $\sin^2\theta_{12}$ ($\Delta m^2_{21}$) is $4.1 \%$ ($2.7 \%$).  
The smallest lepton mixing angle $\theta_{13}$ connects the solar and atmospheric 
sectors, and governs the impact of sub-leading three-flavor effects\,\cite{Pascoli:2013wca,
Agarwalla:2013hma,Agarwalla:2014fva}. The present best fit value of this parameter is 
$8.5^\circ$ with a relative 1$\sigma$ uncertainty of $\sim 1.8 \%$\,\cite{Esteban:2016qun,NuFIT}.
As far as the atmospheric mixing angle is concerned, the $3\sigma$ allowed range of $\sin^2\theta_{23}$ 
is 0.4 to 0.63, and a relative $1\sigma$ precision on this parameter is around $7\%$.  This relatively large 
allowed range in $\theta_{23}$ suggests that it can be maximal or non-maximal\footnote{If 
$\theta_{23}\neq 45^\circ$, there can be two possibilities: one $< 45^\circ$, known as lower 
octant (LO), and other $> 45^\circ$, termed as higher octant (HO).}. Recently, the currently 
running accelerator experiment NO$\nu$A has provided a hint of non-maximal $\theta_{23}$ at 
around $2.6 \sigma$ confidence level\,\cite{Adamson:2017qqn}. For $|\Delta m^2_{32}|$, the present 
best fit value is $2.44\times 10^{-3}$ eV$^2$, the $3\sigma$ allowed range is $2.33\times 10^{-3}$ 
eV$^2$ to  $2.55\times 10^{-3}$ eV$^2$, and a relative  1$\sigma$ uncertainty is $1.5 \%$. The 
current oscillation data cannot decide whether this parameter is positive ($\Delta m^2_{32}>0$)  
or negative ($\Delta m^2_{32}<0$). The first possibility gives rise to the neutrino mass pattern:  
$m_3>m_2>m_1$, known as normal hierarchy (NH) and for the second  possibility, we have 
$m_2>m_1>m_3$, labelled as inverted hierarchy (IH). Recently, in Ref.\,\cite{Abe:2017aap}, an analysis 
of the Super-Kamiokande atmospheric neutrino data over a 328 kt$\cdot$yr exposure of the detector 
has been performed. They find a weak preference for NH, disfavoring IH at $93 \%$ C.L. assuming 
the best fit values of the oscillation parameters obtained from their analysis.  The interesting 
complementarity among the accelerator and reactor data has already provided crucial information 
on the $\delta_{\rm CP}$ phase\,\cite{Esteban:2016qun,deSalas:2017kay,Capozzi:2017ipn, NuFIT}. 
A hint in favor of $\delta_{\rm CP}$ around $-90^\circ$ has been emerged from the global fit studies,  
and this indication is getting strengthened as new data are becoming available. 
Also, the values of $\delta_{\rm CP}$ around  $90^\circ$ ($\in 30^\circ$ to $130^\circ$) are 
already disfavored at more than $3\sigma$ confidence level\,\cite{Esteban:2016qun,deSalas:2017kay,
Capozzi:2017ipn, NuFIT}. 

The proposed 50 kt magnetized Iron Calorimeter (ICAL) detector is designed to observe the 
atmospheric neutrinos and antineutrinos separately over a wide range of energies and 
baselines\,\cite{Kumar:2017sdq,INO}. The main aim of this experiment is to examine the Earth 
matter effect\,\cite{Wolfenstein:1977ue,Mikheev:1986gs, Mikheev:1986wj} by studying the energy 
and zenith angle dependence of the atmospheric neutrinos in the multi-GeV range. It will enable 
the ICAL detector to address some of the fundamental issues in neutrino oscillation physics. 
Preliminary studies have already shown that the INO-ICAL experiment has immense potential 
to determine the neutrino mass hierarchy and to improve the precision on atmospheric neutrino 
mixing parameters\,\cite{Ghosh:2012px,Thakore:2013xqa,Devi:2014yaa,Ajmi:2015uda,Kaur:2014rxa,
Mohan:2016gxm,Kumar:2017sdq}. This facility can also offer an unparalleled window to probe  
the new physics beyond the Standard Model (SM)\,\,\cite{Dash:2014sza,Dash:2014fba,Chatterjee:2014oda,
Chatterjee:2014gxa,Choubey:2015xha,Behera:2016kwr,Choubey:2017eyg,Choubey:2017vpr}. In this paper, we investigate in detail the 
possible impacts of non-universal flavor-diagonal neutral current (FDNC) long-range interactions in the 
oscillations of neutrinos and antineutrinos in the context of INO-ICAL experiment. 
These new interactions come into the picture due to flavor-dependent, vector-like, leptonic 
long-range force (LRF), like those mediated by the $L_e-L_\mu$ or $L_e-L_\tau$ gauge boson, which 
is very light and neutral. 

This paper is organized as follows. In section\,\ref{sec:bound},  we discuss about flavor-dependent 
LRF and how it appears from abelian gauged $L_e-L_{\mu,\tau}$ symmetry. We also estimate the strength 
of long-range potential of $V_{e\mu/e\tau}$ symmetry at the Earth surface generated by the 
electrons inside the Sun. We end this section by mentioning the current constraints that we have 
on the effective gauge couplings $\alpha_{e\mu,e\tau}$ of  the $L_e-L_{\mu,\tau}$ symmetries from 
various experiments. In section \ref{sec:framework}, we study in detail how the three-flavor 
oscillation picture gets modified in presence of long-range potential. We present compact 
analytical expressions for the effective oscillations parameters in presence of LRF. Next, 
we show the accuracy of our analytical probability expressions (for $L_e-L_\tau$) by comparing them  
with the exact numerical results. In appendix\,\ref{app-1},  we perform the similar comparison for the 
$L_e-L_\mu$ symmetry. In section\,\ref{sec:osc}, we draw the neutrino oscillograms in ($E_\nu$, $\cos\theta_\nu$) plane for 
$\nu_e\rightarrow\nu_\mu$ and $\nu_\mu\rightarrow\nu_\mu$ oscillation channels in presence of 
$L_e-L_{\mu,\tau}$ symmetry. We mention the important features of ICAL detector in 
section\,\ref{sec:ical}. In section\,\ref{sec:event}, we show the expected event spectra 
in ICAL with and without LRF. Section\,\ref{sec:sim-proc} deals with the simulation procedure 
that we adopt in this work. Next, we derive the expected constraints on  $\alpha_{e\mu,e\tau}$ 
from ICAL in section\,\ref{sec:results}, and discuss few other interesting results. Finally, 
we summarize and draw our conclusions in section\,\ref{sec:conclusion}.

\section{Flavor-Dependent Long-Range Forces}
\label{sec:bound}
One of the possible ways to extend the SM gauge group SU(3)$_C \times$SU(2)$_L\times$U(1)$_Y$
with minimal matter content is by introducing anomaly free U(1) symmetries with the gauge 
quantum number (for vectorial representations)\,\cite{Ma:1997nq,Lee:2010hf}
\begin{equation}
 Q = a_0(B-L) + a_1(L_e-L_\mu) + a_2(L_e-L_\tau) + a_3(L_\mu-L_\tau)\,.
 \label{eq:anamaly-free}
\end{equation}
Here, $B$ and $L$ are baryon and lepton numbers respectively. $L_l$ are 
lepton flavor numbers and $a_i$ with $i=0,1,2,3$ are arbitrary constants. Note that 
the SM remains invariant and renormalizable if we extend its gauge group in the 
above way\,\cite{Langacker:2008yv}. There are three lepton 
flavor combinations: i) $L_e-L_\mu$ ($a_1 =1$, $a_{0,2,3}=0$), ii) $L_e-L_\tau$ 
($a_2 =1$, $a_{0,1,3}=0$), and iii) $L_\mu-L_\tau$ ($a_3 =1$, $a_{0,1,2}=0$), which can be 
gauged in an anomaly free way with the particle content of the SM\,\cite{Foot:1990mn,
Foot:1990uf,He:1991qd,Foot:1994vd}. In this paper, we concentrate on $L_e-L_{\mu,\tau}$ 
symmetries and the implications of $L_\mu-L_\tau$ symmetry in neutrino oscillation
will be discussed elsewhere. Over the last two decades, it has been confirmed that neutrinos 
do oscillate from one flavor to another, which requires that they should have non-degenerate masses
and mix among each other\,\cite{Olive:2016xmw}. To make it happen, 
the above mentioned U(1) gauge symmetries have to be broken in Nature\,\cite{Joshipura:2003jh,
PhysRevD.75.093005}. 
It is quite obvious that the resultant gauge boson should couple to matter very weakly 
to escape direct detection. On top of it, if the extra gauge boson associated with this 
abelian symmetry is very light, then it can give rise to long-range force having terrestrial 
range (greater than or equal to the Sun-Earth distance) and without introducing extremely 
low mass scales\,\cite{Joshipura:2003jh,Grifols:2003gy,Chatterjee:2015gta}. Interestingly, 
this LRF depends on the leptonic content and the mass of an object. Therefore it violates the 
universality of free fall which can be tested in the classic lunar ranging\,\cite{Williams:1995nq,
Williams:2004qba}, and E\"{o}tv\"{o}s 
type gravity experiments\,\cite{Adelberger:2003zx,Dolgov:1999gk}. Lee and Yang gave this idea long 
back in Ref.\,\cite{Lee:1955vk}.  Later, Okun used their idea and gave a $2 \sigma$ bound on $\alpha< 3.4 
\times 10^{−49}$ ($\alpha$ stands for the strength of long-range potential) for a range of the 
Sun-Earth distance or more\,\cite{Okun:1995dn,Okun:1969ey}. 

The coupling of the solar electron to $L_e-L_{\mu/\tau}$ gauge boson leads to a flavor-dependent 
long-range potential for neutrinos\,\cite{Grifols:1993rs,Grifols:1996fk,Horvat:1996mt}, which can 
affect neutrino oscillations\,\cite{Grifols:2003gy,Joshipura:2003jh,PhysRevD.75.093005,GonzalezGarcia:2006vp,
Samanta:2010zh,Chatterjee:2015gta} in spite of such tight constraint 
on $\alpha$ as mentioned above. Here, ($L_e-L_{\mu/\tau}$)-charge of $\nu_e$ is opposite to that of  
$\nu_\mu$ or $\nu_\tau$, which results in new non-universal FDNC interactions of neutrinos. 
These new interactions along with the standard $W$-exchange interactions between ambient 
electrons and propagating $\nu_e$ in matter can alter the ``running'' of oscillation parameters in non-trivial 
fashion\,\cite{Agarwalla:2013tza}. For an example, the electrons inside the Sun can generate a 
flavor-dependent long-range potential $V_{e\mu/e\tau}$ at the Earth surface which has the following 
form\,\cite{Joshipura:2003jh,PhysRevD.75.093005},
\begin{equation}
V_{e\mu/e\tau}(R_{SE})=\alpha_{e\mu/e\tau}\,\frac{N^{\odot}_e}{R_{SE}}
\approx 1.3\,\times\,10^{-11}\,{\rm eV}\,\Big(\frac{\alpha_{e\mu/e\tau}}
{10^{-50}}\Big)\,,
\label{eq:vemutau}
\end{equation} 
where $\alpha_{e\mu/e\tau} = \frac{g^2_{e\mu/e\tau}}{4\pi}$ is the ``fine structure constant"
of the new abelian symmetry and $g_{e\mu/e\tau}$ is the corresponding gauge coupling.
In above equation, $N^{\odot}_e$ denotes the total number of electrons 
($\approx 10^{57}$) in the Sun\,\cite{bahcall:1989} and $R_{SE}$ is the Sun-Earth distance 
$\approx 1.5\times 10^{13} {\rm cm} = 7.6\times 10^{26}$ GeV$^{-1}$. The LRF potential $V_{e\mu/e\tau}$
in Eq.\,\ref{eq:vemutau} comes with a negative sign for antineutrinos and can be probed  
separately in ICAL along with the corresponding potential for neutrinos. The LRF potential 
due to the electrons inside the Earth with the Earth-radius range ($R_E\sim 6400$ km) is roughly 
one order of magnitude smaller as compared to the potential due to the Sun. 
Therefore, we safely neglect the contributions coming from the Earth\,\cite{Joshipura:2003jh,
PhysRevD.75.093005}.  

There are already tight constraints on the effective gauge coupling $\alpha_{e\mu/e\tau}$ of 
$L_e-L_{\mu/\tau}$ abelian symmetry using the data from various neutrino oscillation experiments. 
In \cite{Joshipura:2003jh}, an upper bound of $\alpha_{e\mu}<5.5\times 10^{-52}$  
at $90\%$ C.L. was obtained using the atmospheric neutrino data of the Super-Kamiokande experiment. 
The corresponding limit on $\alpha_{e\tau}$ is $<6.4\times 10^{-52}$ at $90\%$ confidence level.
A global fit of the solar neutrino and KamLAND data in the presence of LRF was performed in 
\cite{PhysRevD.75.093005}. They gave an upper bound of $\alpha_{e\mu}<3.4\times 10^{-53}$ 
at $3\sigma$ C.L. assuming $\theta_{13}=0^{\circ}$. Their limit on $\alpha_{e\tau}$ is 
$<2.5\times 10^{-53}$ at $3\sigma$. In \cite{GonzalezGarcia:2006vp}, the authors performed a similar analysis 
to derive the limits on LRF mediated by vector and non-vector (scalar or tensor) neutral bosons assuming 
one mass scale dominance. A preliminary study to constrain the LRF parameters in the context ICAL 
detector was carried out in \cite{Samanta:2010zh}. Using an exposure of one Mton$\cdot$yr and 
considering only the muon momentum as observable, an expected upper bound of 
$\alpha_{e\mu/e\tau}\lesssim 1.65\times 10^{-53}$ at $3\sigma$ was obtained for ICAL.

\section{Three-Flavor Neutrino Oscillation with Long-Range Forces}
\label{sec:framework}
In this section, we discuss how the flavor-dependent long-range potential due to 
the electrons inside the sun modify the oscillation of terrestrial neutrinos. In 
presence of LRF, the effective Hamiltonian (in the flavor basis) for neutrino 
propagation inside the Earth is given by
\begin{equation}
H_f=U\left[
\begin{tabular}{c c c} 0 & 0 & 0\\
                      0 & $\frac{\Delta m^2_{21}}{2E}$ & 0\\
                      0 & 0 & $\frac{\Delta m^2_{31}}{2E}$
\end{tabular}\right]U^\dag\, + \left[
\begin{tabular}{c c c} $V_{CC}$ & 0 & 0\\ 0 & 0 & 0 \\ 0 & 0 & 0
\end{tabular} \right] + \left[
\begin{tabular}{c c c} $\zeta$ & 0 & 0 \\ 0 & $\xi$ & 0\\ 
                        0 & 0 & $\eta$ \\
\end{tabular} \right]\,,
\label{eq:Hf}
\end{equation}    
where $U$ is the vacuum PMNS matrix\,\cite{Pontecorvo:1967fh,Pontecorvo:1957qd,
Maki:1962mu}, $E$ denotes the energy of neutrino, and $V_{CC}$ represents the 
Earth matter potential which can be expressed as 
\begin{equation}
V_{CC} = \sqrt{2}\,G_F\,N_{e} \simeq 7.6\times Y_{e} \times \frac{\rho}{10^{14}\,{\rm g/cm^{3}}}\,\, {\rm eV}\,.
\label{eq:vcc}
\end{equation}
In above,  $G_F$ is the Fermi coupling constant, $N_e$ is the number density of electron inside the Earth, 
$\rho$ stands for matter density, and $Y_{e}\,(\frac{N_e}{N_p+N_n})$ is the relative electron number density.
Here, $N_p$ and $N_n$ are the proton and neutron densities respectively. For an electrically neutral and 
isoscalar medium, $N_e = N_p = N_n$ and therefore, $Y_{e}=0.5$. In  Eq.\,\ref{eq:Hf}, $\zeta$, $\xi$, and 
$\eta$ appear due to the long-range potential. In case of  $L_e-L_\mu$ symmetry, $\zeta=-\xi=V_{e\mu}$ with  
$\eta=0$. On the other hand, if the underline symmetry is $L_e-L_\tau$, then $\zeta=-\eta=V_{e\tau}$ with $\xi=0$.
Here, $V_{e\mu}$ ($V_{e\tau}$) is the LRF potential due to the interactions mediated by neutral gauge boson 
corresponding to $L_e-L_\mu$ ($L_e-L_\tau$) symmetry. 
Since the strength of $V_{e\mu/e\tau}$ (see Eq.\,\ref{eq:vemutau}) does not depend on the Earth matter density, 
hence its value remains same for all the baselines. In case of antineutrino, the sign of $V_{CC}$, $V_{e\mu}$, 
$V_{e\tau}$, and $\delta_{CP}$ will be reversed. 

It is evident from Eq.\,\ref{eq:Hf} that if the strength of $V_{e\mu/e\tau}$ is comparable to 
$\Delta m^2_{31}/2E$ and $V_{CC}$, then LRF would certainly affect the neutrino propagation. 
Now, let us consider some benchmark choices of energies ($E$) and baselines ($L$) for which 
the above mentioned quantities are comparable in the context of ICAL detector. This detector is 
quite efficient to detect neutrinos and antineutrinos separately in multi-GeV energy range with 
baselines in the range of 2000 to 8000 km where we have substantial Earth matter effect. Therefore, 
in table\,\ref{tab:comp-values}, we show the comparison for three choices of $E$ and $L$: 
(2 GeV, 2000 km), (5 GeV, 5000 km), and (15 GeV, 8000 km). Using Eq.\,\ref{eq:vcc}, we estimate the 
size of $V_{CC}$ for these three baselines for which the line-averaged  constant Earth matter densities 
($\rho$) based on the PREM\,\cite{PREM:1981} profile are 3.46 g/cm$^3$, 3.9 g/cm$^3$, and 4.26 g/cm$^{3}$ 
respectively. From Eq.\,\ref{eq:vemutau}, we obtain the values of $V_{e\mu/e\tau}$ for two benchmark 
choices of  $\alpha_{e\mu/e\tau}$: $10^{-52}$  and  $5 \times 10^{-52}$ (see last column of 
table\,\ref{tab:comp-values}). We compute the value of $\Delta m^2_{31}/2E$  assuming the best fit value 
of $\Delta m^2_{31}\, =\, 2.524\times10^{-3}\, {\rm eV^2}$\,\cite{Esteban:2016qun}. Table\,\ref{tab:comp-values} 
shows that the quantities $\Delta m^2_{31}/2E$, $V_{CC}$, and $V_{e\mu/e\tau}$ are of comparable strengths 
for our benchmark choices of $E$, $L$, and $\alpha_{e\mu/e\tau}$. It suggests that they can interfere 
with each other to alter the oscillation probabilities significantly. Next, we study the ``running'' of  
oscillation parameters in matter in presence of LRF potential.
  
\begin{table}[htb!]
\begin{center}
\begin{tabular}{|c|c|c|c|c|c|}
\hline
\hline
$L$ (km)  & \multirow{2}{*}{$E$ (GeV)} & \multirow{2}{*}{$\frac{\Delta m^2_{31}}{2E}$ (eV)} 
& \multirow{2}{*}{$V_{CC}$ (eV)}  & \multicolumn{2}{c|}{$V_{e\mu/e\tau}$ (eV)}\\
\cline{5-6}
 \,\,($\cos\theta_\nu$)\,\,   &    &  &  & $\alpha_{e\mu/e\tau}=10^{-52}$ & $\alpha_{e\mu/e\tau}=5 \times 10^{-52}$ \\
\hline
\makecell{2000\\ ($-0.15$)}& 2  &  $6.3\times 10^{-13}$ & $1.3\times 10^{-13}$ & $1.3\times 10^{-13}$&$6.5\times 10^{-13}$ \\  
\hline
\makecell{5000\\  ($ -0.39$)} & 5  &  $2.5\times 10^{-13}$ & $1.5\times 10^{-13}$ & $1.3\times 10^{-13}$&  $6.5\times 10^{-13}$ \\ 
\hline
\makecell{8000 \\($ -0.63$)} & 15  & $ 0.84 \times 10^{-13}$ & $1.6 \times 10^{-13}$ & $1.3\times 10^{-13}$ & $6.5\times 10^{-13}$\\ 
\hline 
\hline
\end{tabular}
\end{center}
\mycaption{The values of $\Delta m^2_{31}/2E$ (third column), $V_{CC}$ (fourth column), 
and  $V_{e\mu/e\tau}$ (fifth column) for our benchmark choices of $E$, $L$, and 
$\alpha_{e\mu/e\tau}$. We take $\Delta m^2_{31}\, =\, 2.524\times10^{-3}\, \rm{eV^2}$.
Based on the PREM profile, the line-averaged constant Earth matter densities for  2000 km, 
5000 km, and 8000 km baselines are 3.46 g/cm$^3$, 3.9 g/cm$^3$, and 4.26 g/cm$^{3}$ respectively.
The parameter $\theta_\nu$ is the zenith angle for a given baseline.}
\label{tab:comp-values}
\end{table}

\subsection{``Running'' of Oscillation Parameters}
\label{subsec:running-exp}
The approximate analytical expressions for the effective mass-squared differences 
and mixing angles in presence of $V_{CC}$ and $V_{e\mu}$ (due to $L_e-L_\mu$ symmetry)
have been given in Ref.\,\cite{Chatterjee:2015gta}. In this paper, we derive the 
analytical expressions for $L_e-L_\tau$ symmetry. Assuming $\delta_{\rm CP}=0^{\circ}$, 
the effective Hamiltonian can be written as 
\begin{equation}
 H_f\,=\,R_{23}(\theta_{23})\,R_{13}(\theta_{13})\,R_{12}(\theta_{12})\,
 H_0\,R_{12}^T(\theta_{12})\,R_{13}^T(\theta_{13})\,R_{23}^T(\theta_{23})
 \,+\,V\,,
 \label{eq:hf-expand}
\end{equation}
where for the PMNS matrix ($U$), we follow the CKM parameterization\,\cite{Olive:2016xmw}.   
In the above equation, $H_0={\rm{Diag}}(0,\Delta_{21},\Delta_{31})$ with 
$\Delta_{21}\equiv\Delta m^2_{21}/2E$ and $\Delta_{31}\equiv\Delta m^2_{31}/2E$.
For  $L_e-L_\tau$ symmetry, $V={\rm{Diag}}(V_{CC}+V_{e\tau},0,-V_{e\tau})$.
Considering maximal mixing for $\theta_{23}$ ($=45^{\circ}$), we rewrite $H_f$ 
in the following way
\begin{equation}
 H_f= \Delta_{31} \left(\begin{tabular}{c c c} $b_{11}$ & $b_{12}$ & $b_{13}$\\
			   $b_{12}$ & $b_{22}$ & $b_{23}$\\
			  $b_{13}$ & $b_{23}$ & $b_{33}$ 
     \end{tabular}\right)\,,
     \label{eq:h_f}
\end{equation}
where 
\begin{equation}
 b_{11} = A\,+\,W\,+\sin^{2}\theta_{13}
 \,+\,\alpha \sin^{2}\theta_{12} \cos^2\theta_{13}\,,
 \label{eq:b11}
\end{equation}
\begin{equation}
 b_{12} = \frac{1}{\sqrt{2}} \left[ \cos\theta_{13}( \alpha\cos\theta_{12}
 \sin\theta_{12}+\sin\theta_{13}-\alpha\sin^2\theta_{12}\sin\theta_{13})\right]\,,
 \label{eq:b12}
\end{equation}
\begin{equation}
 b_{13} = \frac{1}{\sqrt{2}} \left[ \cos\theta_{13}( -\alpha\cos\theta_{12}
 \sin\theta_{12}+\sin\theta_{13}-\alpha\sin^2\theta_{12}\sin\theta_{13})\right]\,,
 \label{eq:b13}
\end{equation}
\begin{equation}
 b_{22} = \frac{1}{2} \left[\cos^2\theta_{13}\,+\,\alpha\cos^2\theta_{12}\,
 -\,\alpha\sin 2\theta_{12}\sin\theta_{13}\,+\,\alpha\sin^2\theta_{12}\sin^2\theta_{13}\right]\,,
 \label{eq:b22}
\end{equation}
\begin{equation}
 b_{23} = \frac{1}{2} \left[\cos^2\theta_{13}\,-\,\alpha\cos^2\theta_{12}\,
 +\,\alpha\sin^2\theta_{12}\sin^2\theta_{13}\right]\,,
 \label{eq:b23}
\end{equation}
\begin{equation}
 b_{33} = \frac{1}{2}\, \left[\cos^2\theta_{13}\,+\,\alpha\cos^2\theta_{12}\,
 +\,\alpha \sin 2\theta_{12} \sin\theta_{13}\,+\,\alpha \sin^2\theta_{12} \sin^2\theta_{13}\,-\,2W \right]\,.
 \label{eq:b33}
\end{equation}
In the above equations, the terms A, $W$, and $\alpha$ are defined as  
\begin{equation}
 A \equiv \frac{V_{CC}}{\Delta_{31}} = \frac{2 E V_{CC}}{\Delta m^2_{31}},\,
 W \equiv \frac{V_{e\tau}}{\Delta_{31}} = \frac{2 E  V_{e\tau}}{\Delta m^2_{31}},\, 
 {\rm and} \,\, \alpha \equiv \frac{\Delta m^2_{21}}{\Delta m^2_{31}}\,.
 \label{eq:awalpha}
\end{equation} 
The following unitary matrix ${\tilde{U}}$  can almost diagonalize the effective 
Hamiltonian ($H_f$): 
\begin{equation}
{\tilde{U}}\equiv R_{23}(\theta^m_{23})\,R_{13}(\theta^m_{13})
\,R_{12}(\theta^m_{12})\,,
\label{eq:utilde}
\end{equation}
such that  
\begin{equation}
 {\tilde{U}}^T\,H_f\,\tilde{U} \simeq {\rm{Diag}}(m^2_{1,m}/2E,
 m^2_{2,m}/2E,m^2_{3,m}/2E)\,.
 \label{eq:utilde-hf}
\end{equation}
In the above equation, we neglect the off-diagonal terms which are small. 
Diagonalizing the (2, 3) block of $H_f$, we get the following 
expression for $\theta^m_{23}$ 
\begin{equation}
 \tan 2 \theta^m_{23} = \frac{\cos^2\theta_{13}\,-\,\alpha\cos^2\theta_{12}\,
 +\,\alpha\sin^2\theta_{12}\sin^2\theta_{13}}{-W\,
 +\,\alpha\sin 2\theta_{12}\sin\theta_{13}}.
 \label{eq:the23m}
\end{equation}
We can obtain the expressions for $\theta^m_{13}$ and $\theta^m_{12}$ 
by diagonalizing the (1,3) and (1,2) blocks subsequently. These 
effective mixing angles can be written in following way 
\begin{equation}
 \tan 2 \theta^m_{13} = \frac{\sin 2\theta_{13}(1\,-\,\alpha
 \sin^2\theta_{12})(\cos\theta^m_{23}\,+\,\sin\theta^m_{23})\,
 -\,\alpha\sin 2\theta_{12}\cos\theta_{13}(\cos\theta^m_{23}\,
 -\,\sin\theta^m_{23})} {\sqrt{2}(\lambda_3\,-\,A\,-\,W
 \,-\,\sin^2\theta_{13}\,-\,\alpha\sin^2\theta_{12}\cos^2\theta_{13})}
\label{eq:the13m}
 \end{equation}
 and
\begin{align}
 &\tan 2 \theta^m_{12} = \nonumber\\&\frac{\cos\theta^m_{13}[\sin 2\theta_{13}
 (1\,-\,\alpha\sin^2\theta_{12})(\cos\theta^m_{23}\,
 -\,\sin\theta^m_{23})\,+\,\alpha\sin 2\theta_{12}\cos\theta_{13}
 (\cos\theta^m_{23}\,+\,\sin\theta^m_{23})}{\sqrt{2}
 (\lambda_2\,-\,\lambda_1)}
 \label{eq:the12m}\,.
\end{align}
In the above expressions, $\lambda_3$, $\lambda_2$, and $\lambda_1$
take the following forms 
\begin{equation}
 \lambda_3 = \frac{1}{2}\bigg[\cos^2\theta_{13}\,+\,
 \alpha\cos^2\theta_{12}\,+\,\alpha\sin^2\theta_{12}\sin^2\theta_{13}
 \,-W + \frac{(\alpha\sin 2\theta_{12}\sin\theta_{13}-W)}
 {\cos2\theta^m_{23}}\bigg]\,,
 \label{eq:lambda3}
\end{equation}
\begin{equation}
 \lambda_2 = \frac{1}{2}\bigg[\cos^2\theta_{13}\,+\,\alpha\cos^2\theta_{12}
 \,+\,\alpha\sin^2\theta_{12}\sin^2\theta_{13}\,-W - 
 \frac{(\alpha\sin 2\theta_{12}\sin\theta_{13}-W)}
 {\cos2\theta^m_{23}}\bigg]\,,
 \label{eq:lambda2}
\end{equation}
and 
\begin{align}
 \lambda_1 = \frac{1}{2}\bigg[\big(\lambda_3\,+\,A\,+W\,+&
  \sin^2\theta_{13}\,+\,\alpha \sin^2\theta_{12} \cos^2\theta_{13}\big)\,
  \nonumber\\&-\frac{(\lambda_3\,-\,A\,-W\,-\,\sin^2\theta_{13}\,
 -\,\alpha \sin^2\theta_{12} \cos^2\theta_{13})}{\cos2\theta^m_{13}}\bigg]\,.
\label{eq:lambda1} 
\end{align}
The eigenvalues $m^2_{i,m}/2E$ ($i$ = 1, 2, 3) can be written in following fashion  
\begin{align}
 \frac{m^2_{3,m}}{2E} = \frac{\Delta_{31}}{2}\,\bigg[\lambda_3\,+\,A\,&+
  \,W\,+\,\sin^2\theta_{13}\,+\,\alpha \sin^2\theta_{12} \cos^2\theta_{13}
  \,\nonumber\\&+\,\frac{\lambda_3\,-\,A\,-W\,
  -\,\sin^2\theta_{13}\,-\,\alpha\sin^2\theta_{12}
  \cos^2\theta_{13}}{\cos2\theta^m_{13}}\bigg]\,,
  \label{eq:m3sq}
\end{align}
\begin{equation}
 \frac{m^2_{2,m}}{2E} = \frac{\Delta_{31}}{2}\left[\lambda_1\,+\,\lambda_2\,
 -\,\frac{\lambda_1\,-\,\lambda_2}{\cos 2\theta^m_{12}}\right]\,,
\label{eq:m2sq}
 \end{equation}
and 
\begin{equation}
\frac{m^2_{1,m}}{2E} = \frac{\Delta_{31}}{2}\left[\lambda_1\,+\,\lambda_2\,
+\,\frac{\lambda_1\,-\,\lambda_2}{\cos 2\theta^m_{12}}\right]\,.
\label{eq:m1sq}
\end{equation}

\begin{figure}[htb!]
 \subfigure[]{\includegraphics[width=7.5cm]{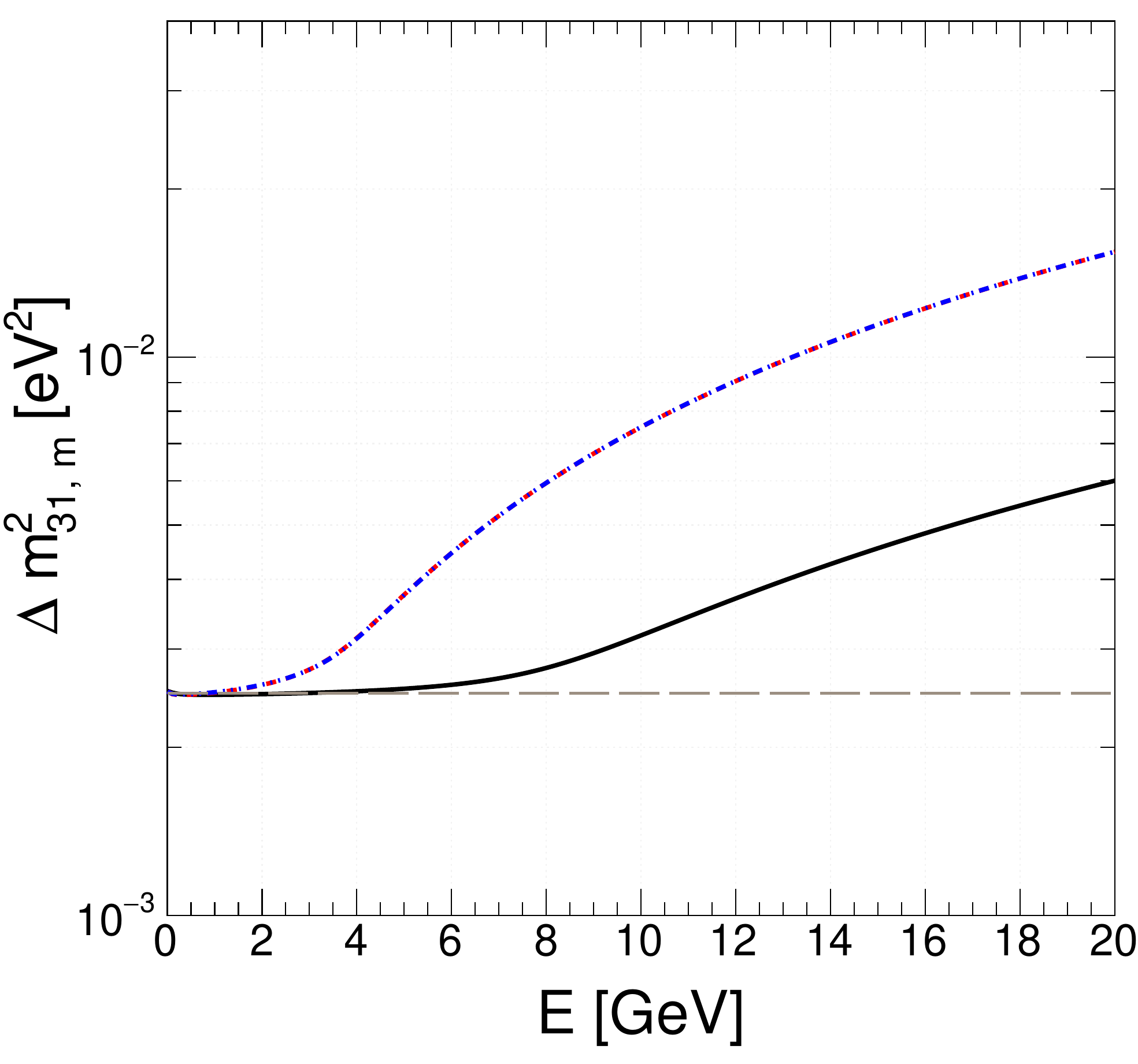}}
 \subfigure[]{\includegraphics[width=7.5cm]{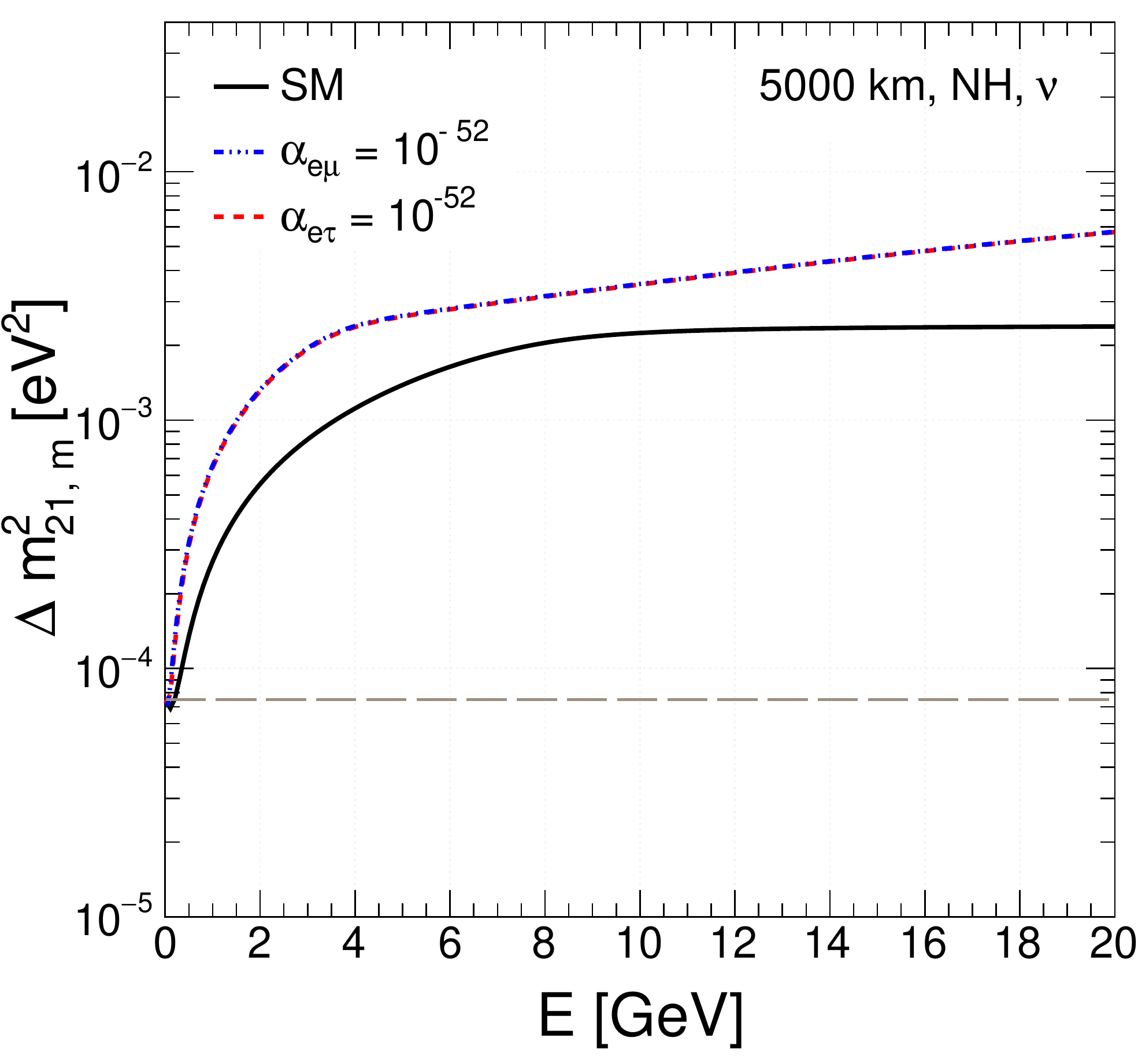}}
\mycaption{ The variations in the $\Delta m^2_{31,m}$ ($\equiv m^2_{3,m} - m^2_{1,m}$, left panel) and $\Delta m^2_{21,m}$ 
($\equiv m^2_{2,m} - m^2_{1,m}$, right panel) with the neutrino energy $E$ in presence of $V_{CC}$ and $V_{e\mu/e\tau}$
for $L$=5000 km and NH. We give plots for three different cases: i) $\alpha_{e\mu} =  
\alpha_{e\tau} =0$ (the SM case, black solid line), ii)  $\alpha_{e\mu} = 10^{-52}$, $\alpha_{e\tau} =0$ (blue dash-dotted line), and 
iii) $\alpha_{e\mu} = 0$, $\alpha_{e\tau} =10^{-52}$ (red dashed line). }
\label{fig:msq-vary}
\end{figure}

To observe the ``running'' of oscillation parameters in presence of $V_{CC}$ and $V_{e\mu/e\tau}$, 
we take the following benchmark values of vacuum oscillation parameters: $\sin^2 \theta_{23} = 0.5$,  
$\,\sin^2 2\theta_{13}= 0.0847$,  $\sin^2 \theta_{12}\, =\, 0.306$, $\Delta m^2_{21}\, =\, 7.5\times10^{-5}\,
\rm{eV^2}$, $\Delta m^2_{31}\, =\, 2.524\times10^{-3}\, \rm{eV^2}$.
\begin{figure}
 \subfigure[]{\includegraphics[width=4.94cm]{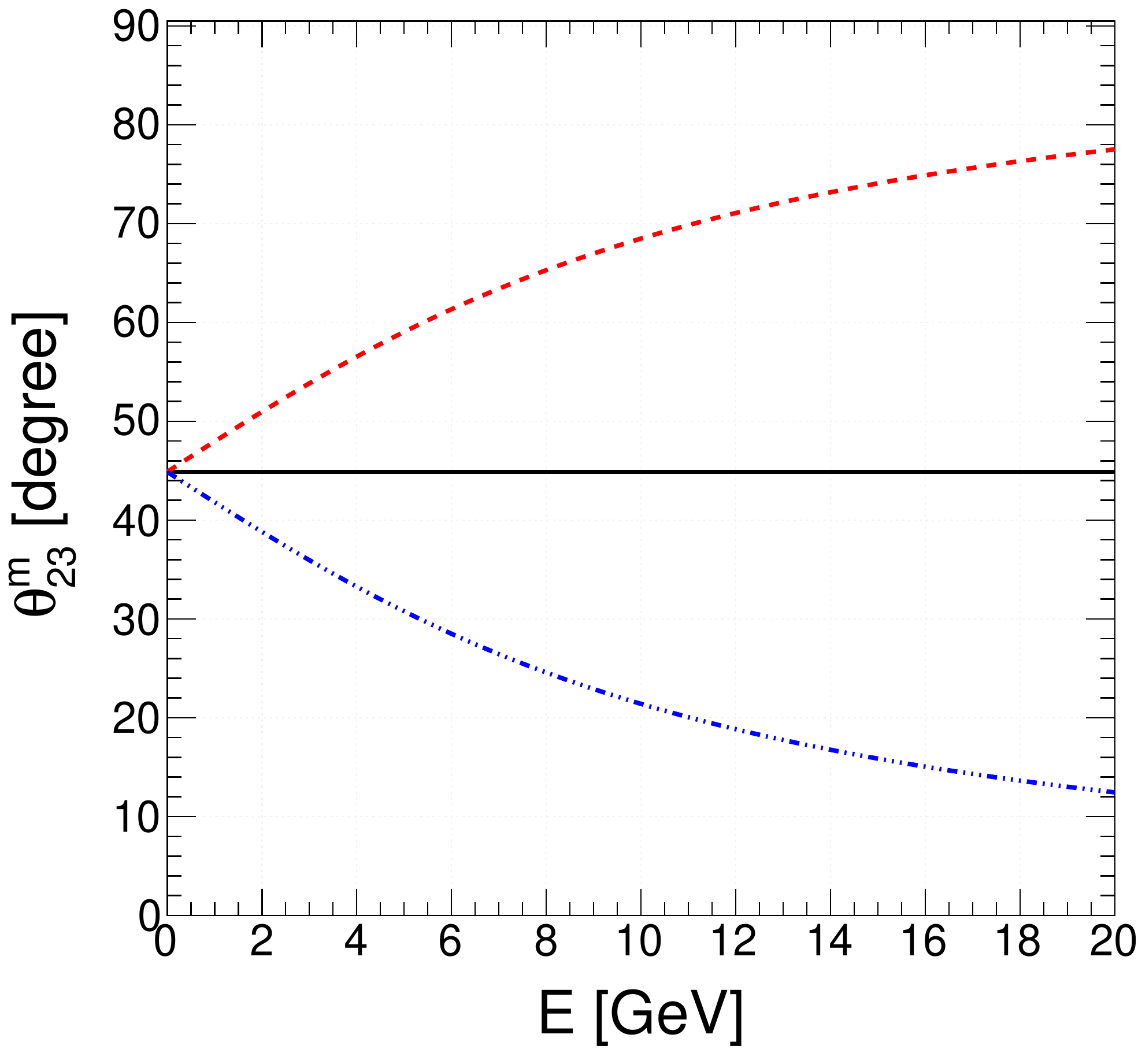}}
 \subfigure[]{\includegraphics[width=4.94cm]{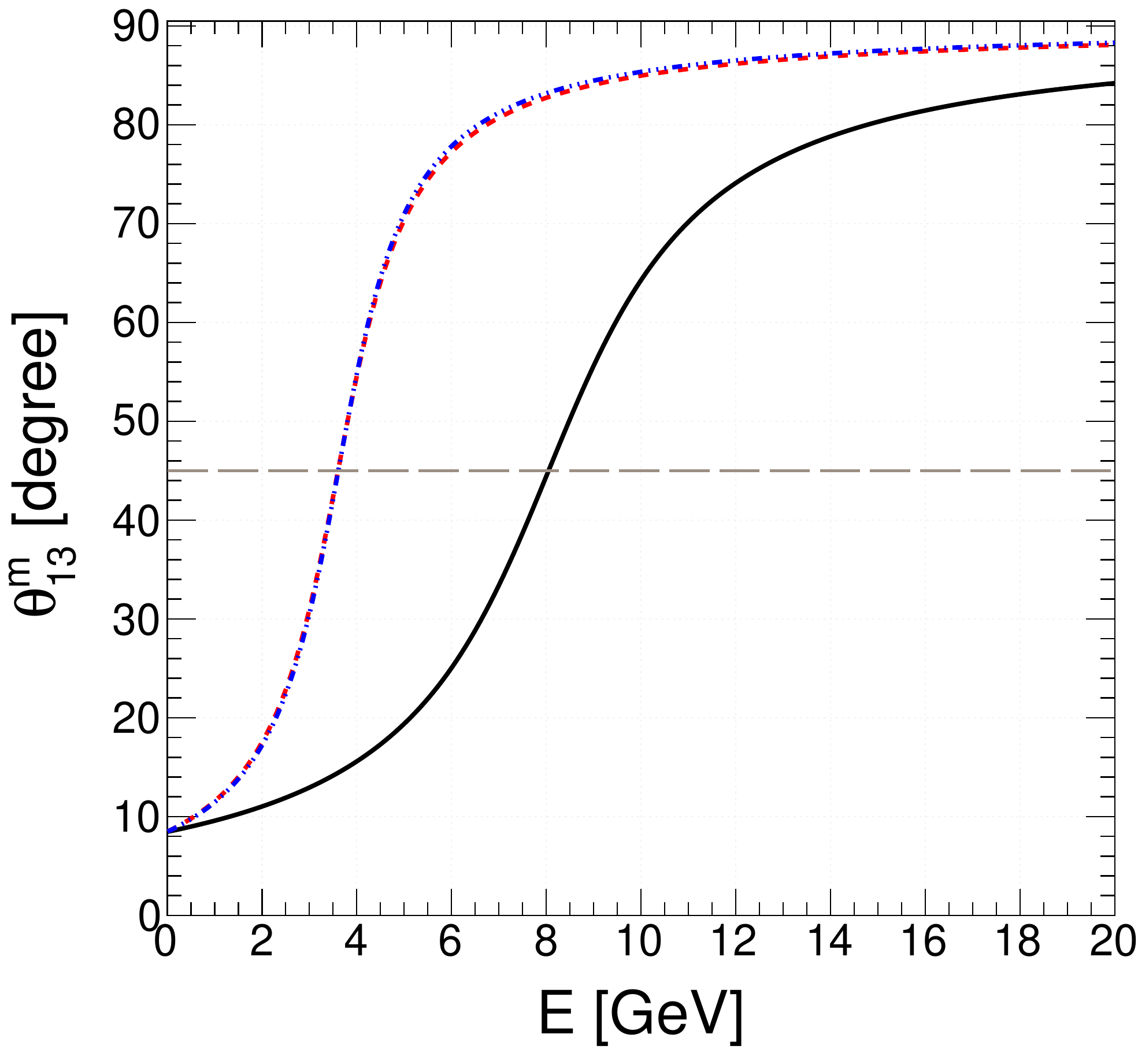}}
 \subfigure[]{\includegraphics[width=4.94cm]{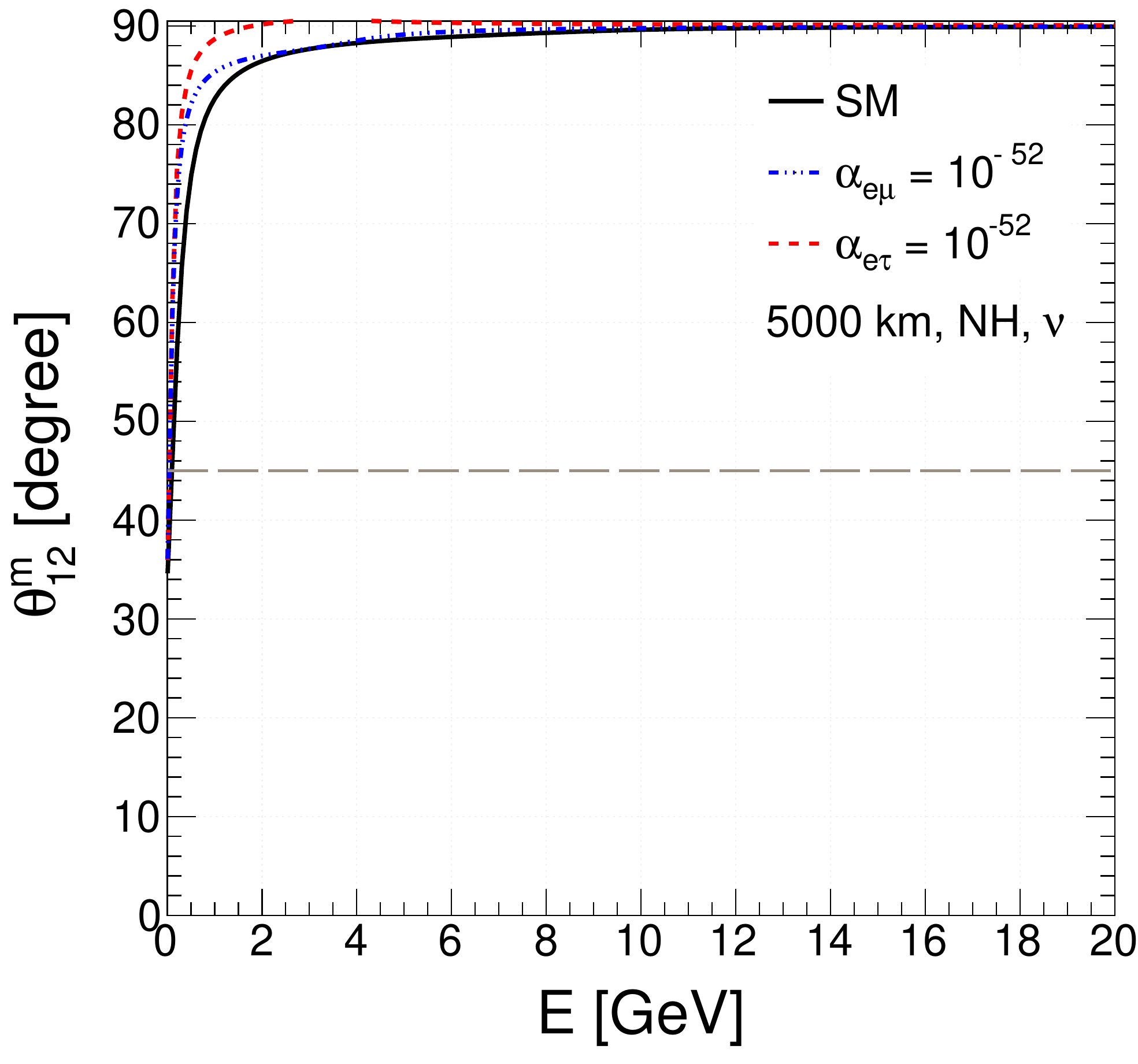}}  
\mycaption{The variations in the effective mixing angles with the neutrino energy 
$E$ in the presence of $V_{CC}$ and $V_{e\mu/e\tau}$. The left, middle, and 
right panels depict the ``running'' of $\theta^{m}_{23}$, $\theta^{m}_{13}$, and 
$\theta^{m}_{12}$ respectively for $L$= 5000 km and NH. In each panel, the black solid line 
is for the SM case, whereas the blue dash-dotted and red dashed lines are for 
$\alpha_{e\mu} = 10^{-52}$ and $\alpha_{e\tau} = 10^{-52}$ respectively.}
\label{fig:th-vary}
\end{figure}
In Fig.\,\ref{fig:th-vary}, we plot the ``running'' of $\theta^{m}_{23}$ (left panel), 
$\theta^{m}_{13}$ (middle panel), and $\theta^{m}_{12}$ (right panel) as functions
of the neutrino energy $E$. These plots are for neutrino with $L= 5000$ km and NH.  
In each panel, we draw the curves for the following three cases\footnote{In case of non-zero 
$\alpha_{e\tau}$, we use Eq.\,\ref{eq:the23m}, Eq.\,\ref{eq:the13m}, and Eq.\,\ref{eq:the12m}. 
For non-zero  $\alpha_{e\mu}$, we take the help of Eq.\,3.16, Eq.\,3.17, 
and Eq.\,3.18 as given in Ref.\,\cite{Chatterjee:2015gta}. }: i) $\alpha_{e\mu} =  
\alpha_{e\tau} =0$ (the SM case), ii)  $\alpha_{e\mu} = 10^{-52}$, $\alpha_{e\tau} =0$ 
iii) $\alpha_{e\mu} = 0$, $\alpha_{e\tau} =10^{-52}$. We repeat the same exercise 
for the effective mass-squared differences\footnote{For non-zero $\alpha_{e\tau}$, we obtain the running of 
$\Delta m^2_{31,m}$ and $\Delta m^2_{21,m}$ using Eq.\,\ref{eq:m3sq}, Eq.\,\ref{eq:m2sq}, 
and Eq.\,\ref{eq:m1sq}. For finite $\alpha_{e\mu}$, we derive the same using Eq.\,3.22, Eq.\,3.23, and Eq.\,3.24 
as given in Ref.\,\cite{Chatterjee:2015gta}.} in Fig.\,\ref{fig:msq-vary}. From the extreme right panel 
of Fig.\,\ref{fig:th-vary}, we can see that $\theta^{m}_{12}$ approaches to $90^\circ$ very rapidly 
as we increase $E$. This behavior is true for the SM case and as well as for non-zero 
$\alpha_{e\mu/e\tau}$, but it is not true for $\theta^{m}_{23}$ and $\theta^{m}_{13}$.
The long-range potential $V_{e\mu/e\tau}$ affect the ``running'' of $\theta^m_{23}$ significantly as can be seen 
from the extreme left panel of Fig.\,\ref{fig:th-vary}. As we approach to higher energies, $\theta^{m}_{23}$ 
deviates from the maximal mixing and its value decreases (increases) very sharply for non-zero $\alpha_{e\mu}$ 
($\alpha_{e\tau}$). This opposite behavior in ``running'' of $\theta^{m}_{23}$ for finite $\alpha_{e\mu}$ and 
$\alpha_{e\tau}$ affect the oscillation probabilities in different manner, which we discuss in next subsection. 
Note that $\theta^{m}_{23}$ is independent of $V_{CC}$ (see Eq.\,\ref{eq:the23m}). Therefore, its value 
remains same for all the baselines and same is true for the SM case as well as for non-zero 
$\alpha_{e\mu/e\tau}$. In case of $\theta^{m}_{13}$ (see middle panel of Fig.\,\ref{fig:th-vary}), the impact of 
$V_{e\mu}$ and $V_{e\tau}$ are same and its ``running`` is quite different as compared to $\theta^{m}_{23}$.
Assuming NH, as we go to higher energies, $\theta^{m}_{13}$ quickly reaches to maximal mixing (resonance point)
for both the symmetries as compared to the SM case. Finally, it approaches toward $90^\circ$ as we further increase 
the energy. For $\alpha_{e\mu/e\tau}=10^{-52}$, the resonance occurs around 3.5 GeV for 5000 km baseline.     
An analytical expression for the resonance energy can be obtained from Eq.\,\ref{eq:the13m} assuming 
$\theta^m_{13}=45^\circ$. In one mass scale dominance approximation ($\Delta m^2_{21}=0,\,i.e.\,\alpha=0$), 
the expression for the resonance energy $E_{\rm res}$ can be obtained from the following: 
\begin{equation}
 \lambda_3\,=\,A\,+\,W
 \,+\,\sin^2\theta_{13}\,.
\label{lambda3-ex2}
\end{equation}
Assuming $\alpha=0$ in Eqs.\,\ref{eq:lambda3} and \ref{eq:the23m}, we get a simplified 
expression of  $\lambda_3$ which appears as 
\begin{equation}
 \lambda_3\,=\,\frac{1}{2}\,[\cos^2\theta_{13} - W + 
 \sqrt{W^2_{e\tau}\,+\,\cos^4\theta_{13}}]\simeq\,\frac{1}{2}
 [2\cos^2\theta_{13}-W],
 \label{lambda3-ex1}
\end{equation}
since at $E_{\rm res}$, the term  $W^2$ is small compared to $\cos^4\theta_{13}$, and we can 
safely neglect it. Comparing Eq.\,\ref{lambda3-ex1} and Eq.\,\ref{lambda3-ex2}, 
we obtain a simple and compact expression for $E_{\rm res}$:  
\begin{equation}
 E_{res} = \frac{\Delta m^2_{31}\,\cos 2\theta_{13}}{2V_{CC}\,
 +3V_{e\tau}}\,.
 \label{eq:eres}
\end{equation}
Note that in the absence of LRF, the above equation boils down to the well-known expression for $E_{\rm res}$ 
in the SM case. Also, we notice that the expression for resonance energy is same for both $L_e-L_\tau$
and $L_e-L_\mu$ symmetries (see Eq.\,3.27 in \cite{Chatterjee:2015gta}). It is evident from 
Eq.\,\ref{eq:eres} that for a fixed baseline, in the 
presence of $V_{e\mu/e\tau}$, the resonance takes place at lower energy as compared to the SM case 
(see middle panel of Fig.\,\ref{fig:th-vary}). 
We observe from both the panels of Fig.\,\ref{fig:msq-vary} that  
in presence of LRF, the variations in $\Delta m^2_{31,m}$ and $\Delta m^2_{21,m}$ with energy are different 
as compared to the SM case. Interesting to note that both $V_{e\mu}$ and $V_{e\tau}$ modify the values of effective 
mass-squared differences in same fashion. In case of $\Delta m^2_{21,m}$ (see right panel of Fig.\,\ref{fig:msq-vary}), 
it increases with energy and can be comparable to the vacuum value of $\Delta m^2_{31}$ at around $E=10$ GeV for both 
the SM and SM\,+\,LRF scenarios.  For  $\Delta m^2_{31,m}$ (see left panel of 
Fig.\,\ref{fig:msq-vary}), the change with energy is very mild in the SM case, but in presence of LRF, 
$\Delta m^2_{31,m}$ gets increased substantially as we approach to higher energies. 
In case of antineutrino, the ``running'' of oscillation parameters can be obtained 
in the similar fashion by just replacing $A \rightarrow -A$ and $W \rightarrow -W$ 
in Eqs.\,\ref{eq:the23m} to \ref{eq:m1sq}. Next, we compare the neutrino and antineutrino 
oscillation probabilities obtained from our analytical expressions with those 
calculated numerically.

\subsection{Comparison between Analytical and Numerical Results}
\label{subsec:com-ana-num}
We obtain the analytical probability expressions in the presence of $V_{CC}$ and $V_{e\mu/e\tau}$ 
by replacing the well known vacuum values of the elements of $U_{\rm PMNS}$ and 
the mass-squared differences $\Delta m^2_{ij}$ with their effective ``running'' values 
as discussed in the previous section.   
\begin{figure}[htb!]
 \subfigure[]{\includegraphics[width=7.5cm]{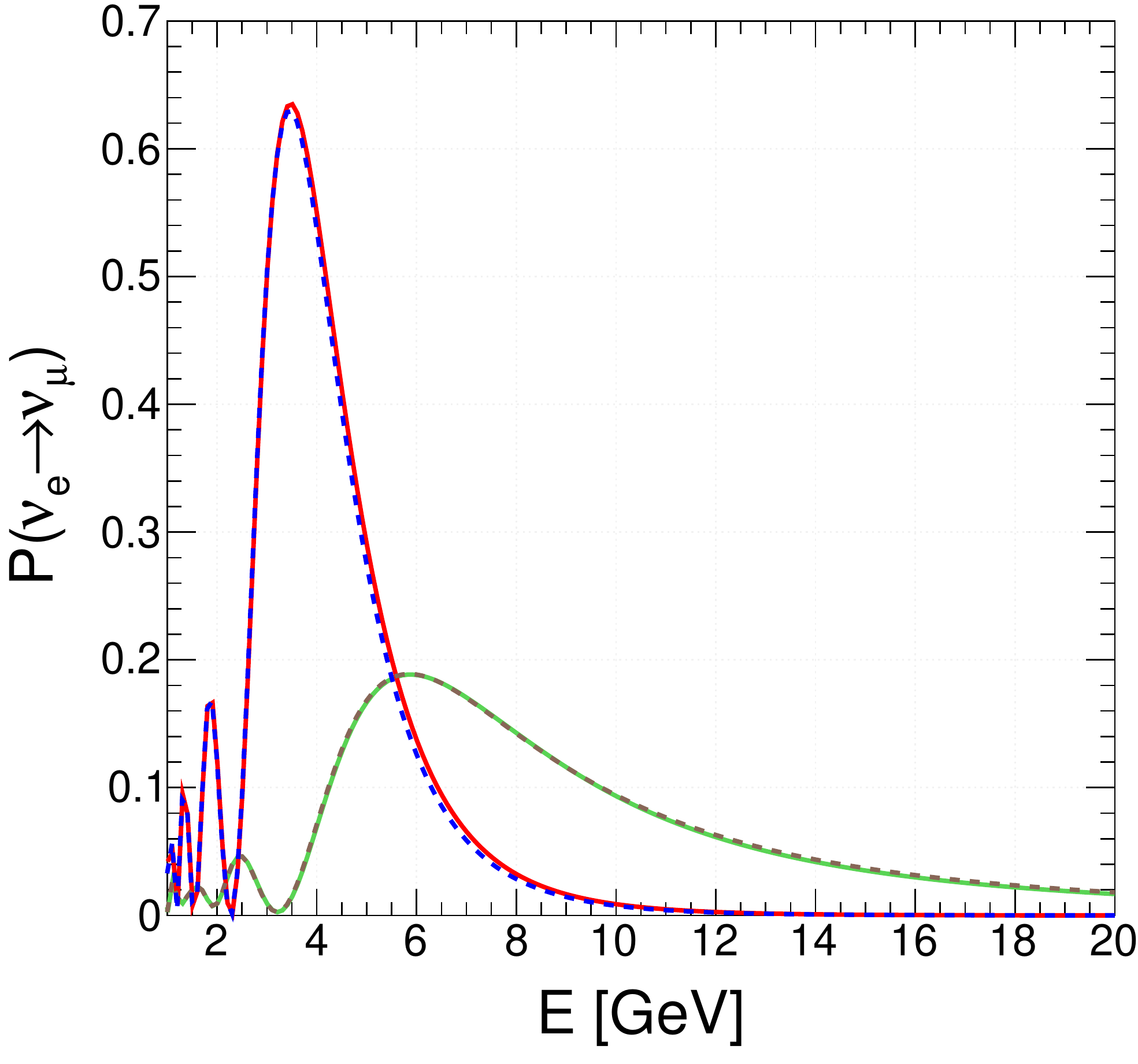}}
 \subfigure[]{\includegraphics[width=7.5cm]{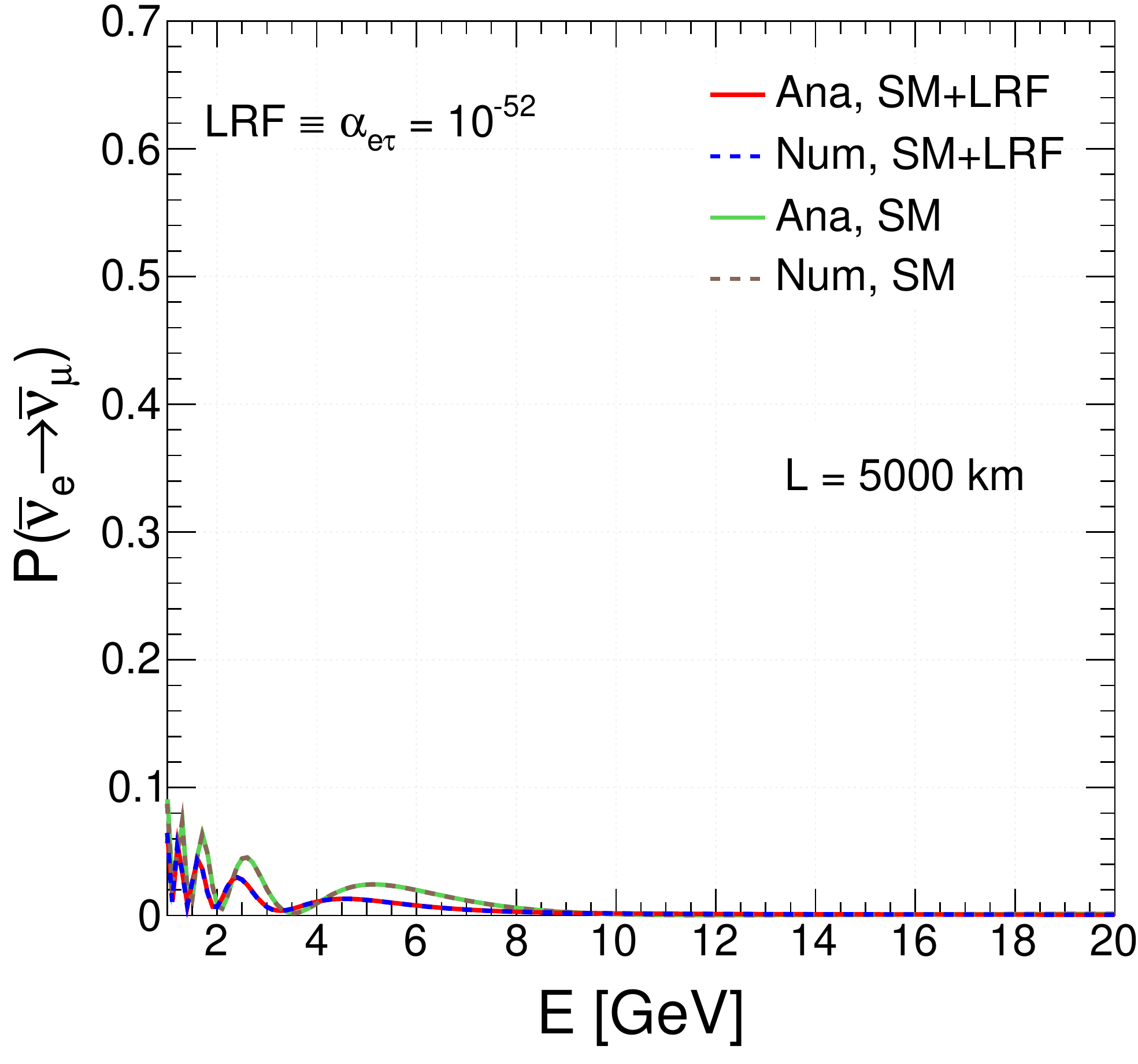}}
 \subfigure[]{\includegraphics[width=7.5cm]{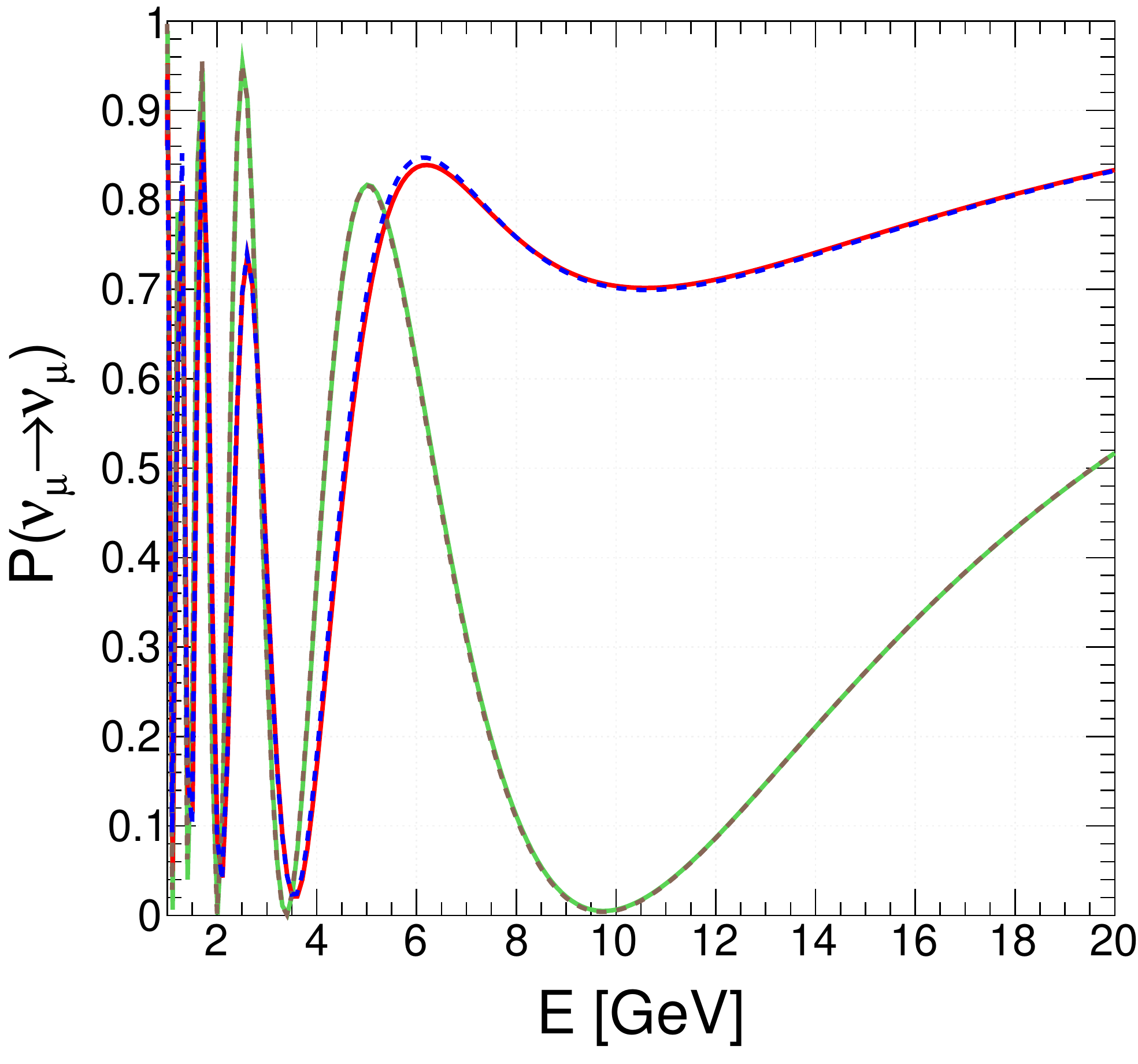}}
 \subfigure[]{\includegraphics[width=7.5cm]{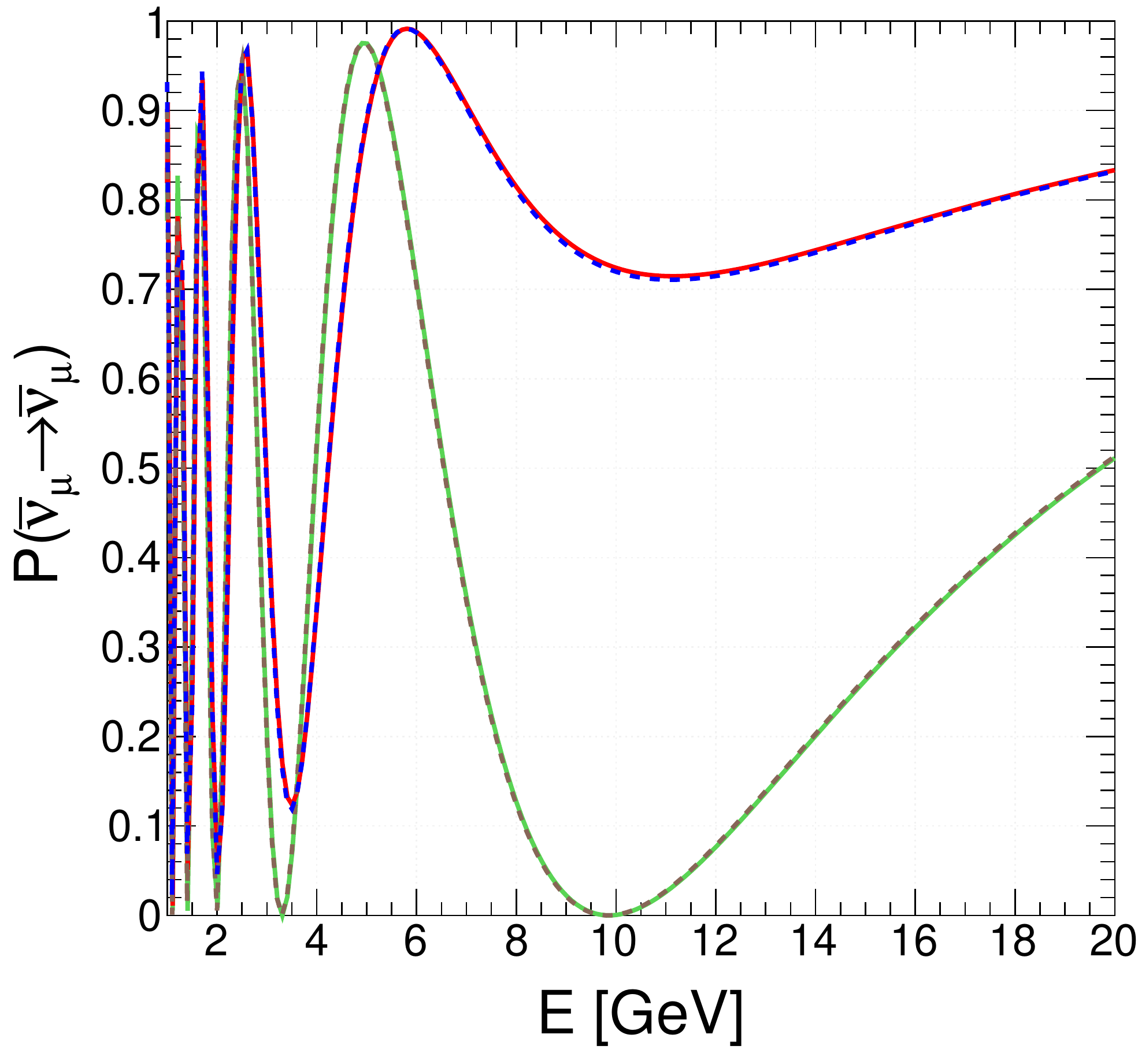}}
\mycaption{ $\nu_e\rightarrow\nu_\mu$ ($\bar\nu_e\rightarrow\bar\nu_\mu$) transition probability for 
5000 km in upper left (right) panel assuming NH. In bottom left (right) panel, we show $\nu_\mu\rightarrow
\nu_\mu$ ($\bar\nu_\mu\rightarrow\bar\nu_\mu$) survival probability. In all the panels, we 
compare our analytical expressions (solid curves) to the exact numerical results (dashed curves) 
for the SM and SM\,+\,LRF cases. For LRF, we consider $\alpha_{e\tau}=10^{-52}$. }
\label{prob-nu-anu-analyt}
\end{figure}
In Fig.\,\ref{prob-nu-anu-analyt}, we show our approximate $\nu_e \rightarrow\nu_\mu$ 
($\bar\nu_e\rightarrow\bar\nu_\mu$) oscillation probabilities in the top left (right) 
panel as a function of $E$ against the exact numerical results considering 
$L = 5000$ km\footnote{For both analytical and numerical calculations, we take the 
line-averaged constant Earth matter density based on the PREM profile\,\cite{PREM:1981}.} 
and NH. We repeat the same for $\nu_\mu \rightarrow\nu_\mu$ ($\bar\nu_\mu \rightarrow \bar\nu_\mu$) 
survival channels  in bottom left (right) panel. We perform these comparisons among 
analytical (solid curves) and numerical (dashed curves) cases for both the SM and SM\,+\,LRF 
scenarios assuming our benchmark choice of $\alpha_{e\tau}=10^{-52}$. For $L_e-L_\mu$ 
symmetry, we perform the similar comparison in Fig.\,\ref{fig:emu-prob-nu-anu-analyt} (see appendix 
\ref{app-1}). 
For the SM case ($\alpha_{e\tau}=0$), our approximate results match exactly with numerically obtained 
probabilities. In the presence of $L_e-L_\tau$ symmetry, we see that our analytical expressions 
work quite well and can produce almost accurate $L/E$ oscillation patterns. 

We can see from the top left panel of Fig.\,\ref{prob-nu-anu-analyt} that for non-zero $\alpha_{e\tau}$,  
the location of the first oscillation maximum shifts toward lower energy (from 5.8 GeV to 3.5 GeV) 
and also its amplitude gets enhanced (from 0.18 to 0.64) for $\nu_e\rightarrow 
\nu_\mu$ transition probability assuming NH. To understand this feature, we can use the 
following simple expression\footnote{We obtain this formula using the general expression as 
given in Eq.\,3.30 in Ref.\,\cite{Chatterjee:2015gta}.} for $P (\nu_e \rightarrow \nu_\mu)$ 
considering $\theta^m_{12}=90^{\circ}$ (see right panel of Fig.\,\ref{fig:th-vary}): 
\begin{equation}
 P_{e\mu} = \sin^2\theta^m_{23}\,\sin^2 2 \theta^m_{13}\,
 \sin^2\frac{\Delta m^2_{32,m}\,L}{4E}\,.
 \label{eq:pemu}
\end{equation}
As can be seen from the previous section, $\theta^m_{23}$ does not ``run'' for the SM case, 
but for non-zero $\alpha_{e\tau}$, it approaches toward $90^\circ$ as we increase $E$. 
As far as $\theta^m_{13}$ is concerned, it quickly reaches to the resonance point 
at a lower energy for non-zero $\alpha_{e\tau}$ as compared to $\alpha_{e\tau}=0$ case. 
Also, $\Delta m^2_{32,m}$ ($\Delta m^2_{31,m} - \Delta m^2_{21,m}$) decreases with energy   
as $\Delta m^2_{21,m}$ increases substantially in comparison to $\Delta m^2_{31,m}$ 
till $E\sim 4$ GeV for 5000 km baseline.    
All these different ``running''  of oscillation parameters are responsible to shift 
the location of first oscillation maximum toward lower energy and also to enhance 
its amplitude. 

In case of $\nu_\mu\rightarrow\nu_\mu$ survival probability ($P_{\mu\mu}$), we can use the 
following simple expression assuming  $\theta^m_{12}=90^\circ$:
\begin{eqnarray}
 P_{\mu\mu} = 1 - \sin^2 2\theta^m_{23}\,\big[\cos^2\theta^m_{13} 
 \sin^2\frac{\Delta m^2_{31,m}\,L}{4E}\,+\,\frac{1}{4}\tan^2\theta^m_{23} 
 \sin^2 2\theta^m_{13} \sin^2\frac{\Delta m^2_{32,m}\,L}{4E}   
	      \nonumber\\ +  \,\sin^2\theta^m_{13} 
	      \sin^2\frac{\Delta m^2_{21,m}\,L}{4E}  \big ]\,.
\label{eq:pmumu}	      
\end{eqnarray}
In the above expression, the term $\sin^2 2\theta^m_{23}$ plays an important role. Now,  
we see from left panel of Fig.\,\ref{fig:th-vary} that as we go to higher energies, $\theta^m_{23}$ 
deviates from the maximal mixing very sharply in presence of LRF. For this reason, 
the value of $\sin^2 2\theta^m_{23}$ gets reduced substantially, which ultimately 
enhances the survival probability for non-zero $\alpha_{e\tau}$ as can be seen from the 
bottom left panel of Fig.\,\ref{prob-nu-anu-analyt}. In the energy range of 6 to 20 GeV,  
we see a substantial enhancement 
in $P_{\mu\mu}$ with non-zero $\alpha_{e\tau}$ as compared to the SM case. 
The same is true for non-zero $\alpha_{e\mu}$  as can be seen from Fig.\,\ref{fig:emu-prob-nu-anu-analyt} 
in appendix\,\ref{app-1}. 
We see a similar increase in case of $\bar\nu_\mu\rightarrow\bar\nu_\mu$ survival 
channel with NH (see bottom right panel of  Fig.\,\ref{prob-nu-anu-analyt}).
We observe this behavior for other baselines as well in  
Figs.\,\ref{fig:osc-pmumu} and \ref{fig:event-letau-lemu}, which we discuss later.


\section{Neutrino Oscillograms in ($E_\nu$, $\cos\theta_\nu$) Plane}
\label{sec:osc}
The atmospheric neutrino experiments deal with a wide range of baselines and energies.
Therefore, it is quite important to see how the long-range forces under discussion 
affect the neutrino oscillation probabilities for all possible choices of baseline 
($\cos\theta_\nu$) and energy ($E_\nu$) which are relevant for the ICAL detector. 
We perform this study by drawing the neutrino oscillograms in ($E_\nu,\,\cos\theta_\nu$) 
plane using the full three-flavor probability expressions with the varying Earth matter 
densities as given in the PREM profile\,\cite{PREM:1981}.   
\subsection{Oscillograms for $\nu_e\rightarrow\nu_\mu$ Appearance Channel}
\begin{figure}[t!]
  \subfigure{\includegraphics[width=0.325\linewidth]{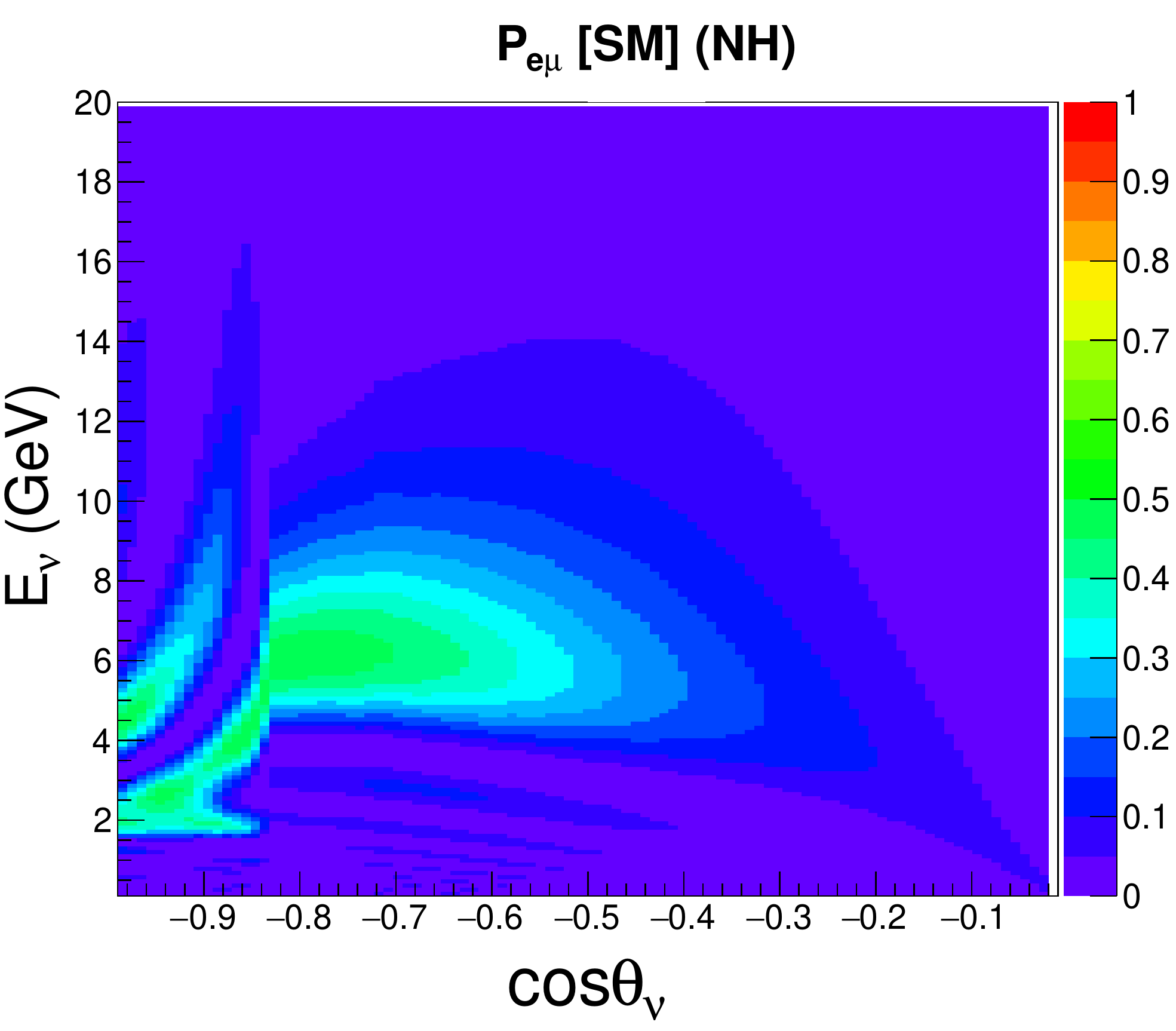}}
\subfigure{\includegraphics[width=0.325\linewidth]{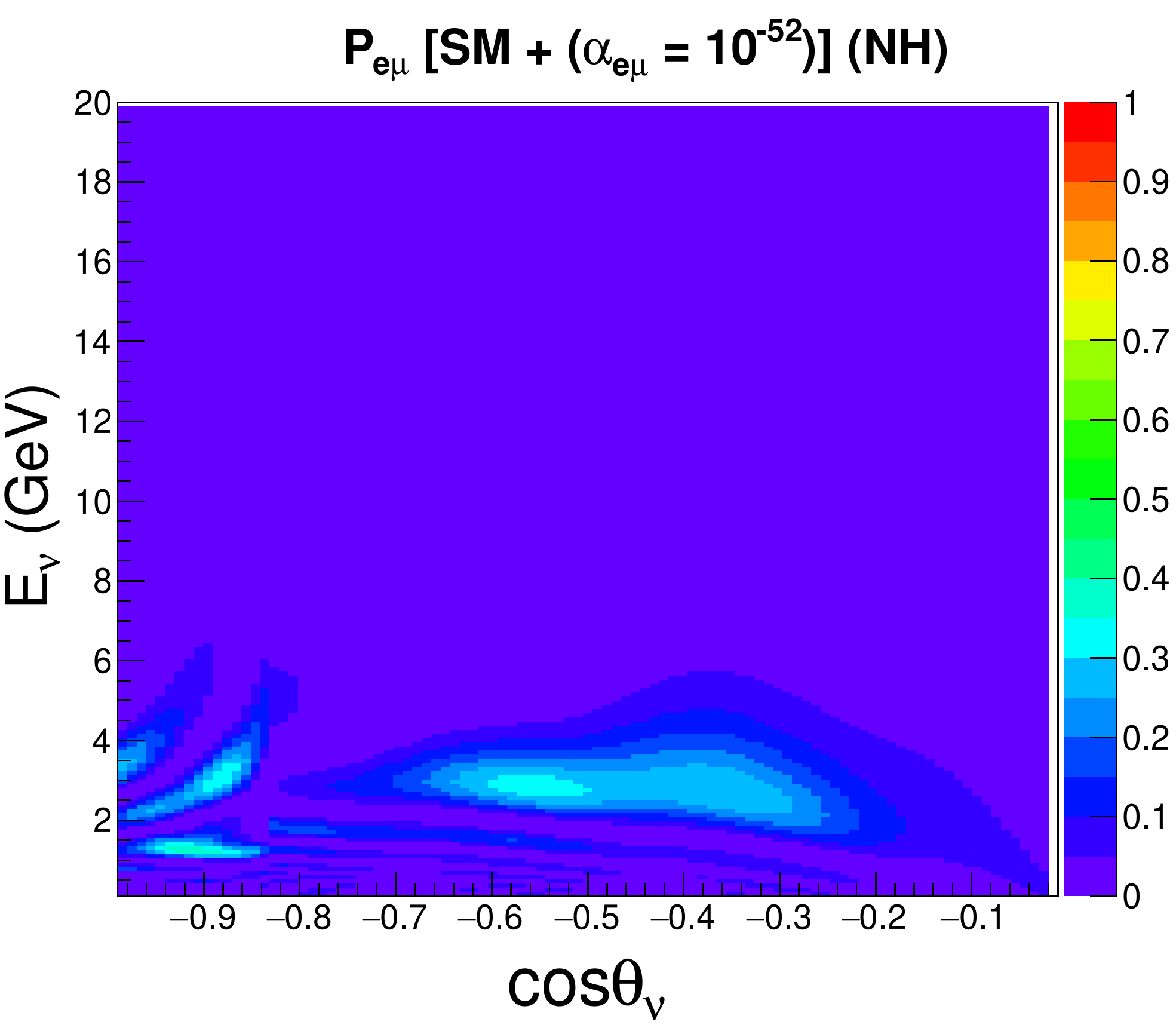}}
 \subfigure{\includegraphics[width=0.325\linewidth]{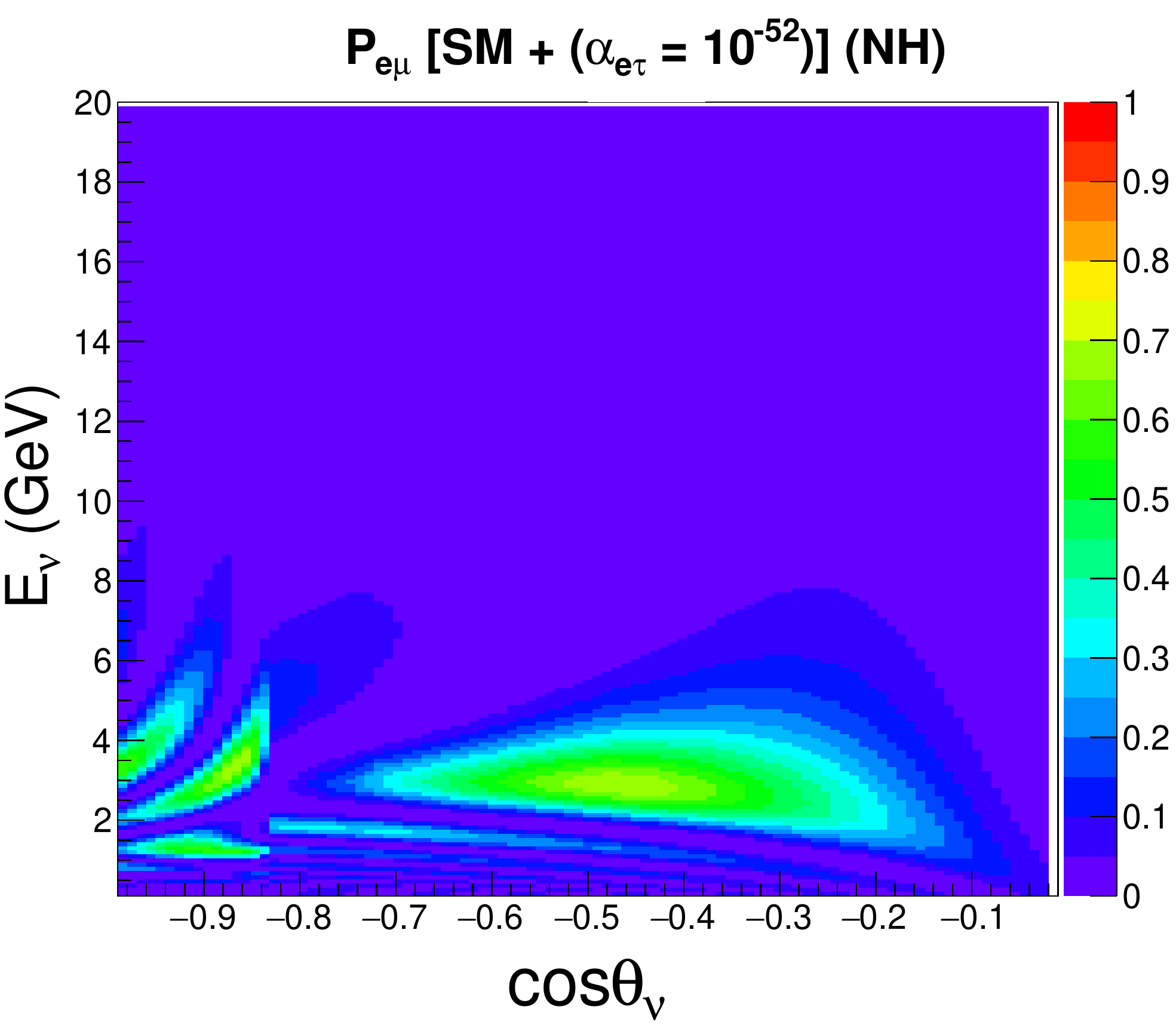}}
 \mycaption{The oscillograms for $\nu_e\rightarrow\nu_\mu$ channel in  $E_\nu$, $\cos\theta_{\nu}$ plane
 for three different scenario: i) $\alpha_{e\mu}=\alpha_{e\tau} =0$ (the SM case, left panel), ii) $\alpha_{e\mu} 
 = 10^{-52}$, $\alpha_{e\tau} =0$ (middle panel), and iii) $\alpha_{e\mu}  = 0$, $\alpha_{e\tau} =10^{-52}$ 
 (right panel). Here, in all the panels, we assume NH. }
 \label{fig:osc-pemu}
 \end{figure}

Fig.\,\ref{fig:osc-pemu} shows the oscillograms for $\nu_e$ to $\nu_\mu$ 
appearance channel in $E_\nu$ and $\cos\theta_\nu$ plane assuming NH. We present 
the oscillograms for three different cases: i) extreme left panel is for the SM case 
($\alpha_{e\mu}=\alpha_{e\tau}=0$), ii) middle panel is for the SM + LRF 
($\alpha_{e\mu}=10^{-52}$), and iii) extreme right panel deals with the SM + LRF 
($\alpha_{e\tau}=10^{-52}$). For the SM case, $\nu_e$ to $\nu_\mu$ 
transition probability attains the maximum value around the resonance region which occurs in 
the range of $E\in$ 4 to 8 GeV and $\cos\theta_\nu\in$ -0.8 to -0.4.  The resonance 
condition in presence of LRF (see Eq.\,\ref{eq:eres}) suggests that $\theta^m_{13}$ can reach 
$45^\circ$ at smaller energies and baselines as compared to the SM case. This feature gets 
reflected in the middle and right panels of Fig.\,\ref{fig:osc-pemu} for non-zero $\alpha_{e\mu}$ and 
$\alpha_{e\tau}$ respectively.  
Fig.\,\ref{fig:osc-pemu} also depicts that the value of $P_{e\mu}$ decreases (increases) as compared to 
the SM case for non-zero $\alpha_{e\mu}$ ($\alpha_{e\tau}$). We can explain this behavior from the 
``running'' of $\theta^m_{23}$ (see left panel of Fig.\,\ref{fig:th-vary}). In presence of $L_e-L_\mu$ 
($L_e-L_\tau$) symmetry, the term $\sin^2\theta^m_{23}$ in Eq.\,\ref{eq:pemu} gets reduced (enhanced) 
as compared to the SM case, which subsequently decreases (increases) the value of $P_{e\mu}$.  

\subsection{Oscillograms for $\nu_\mu\rightarrow\nu_\mu$ Disappearance Channel }
\label{subsec:osc-pmumu}
\begin{figure}[htb!]
 \subfigure{\includegraphics[width=0.32\linewidth]{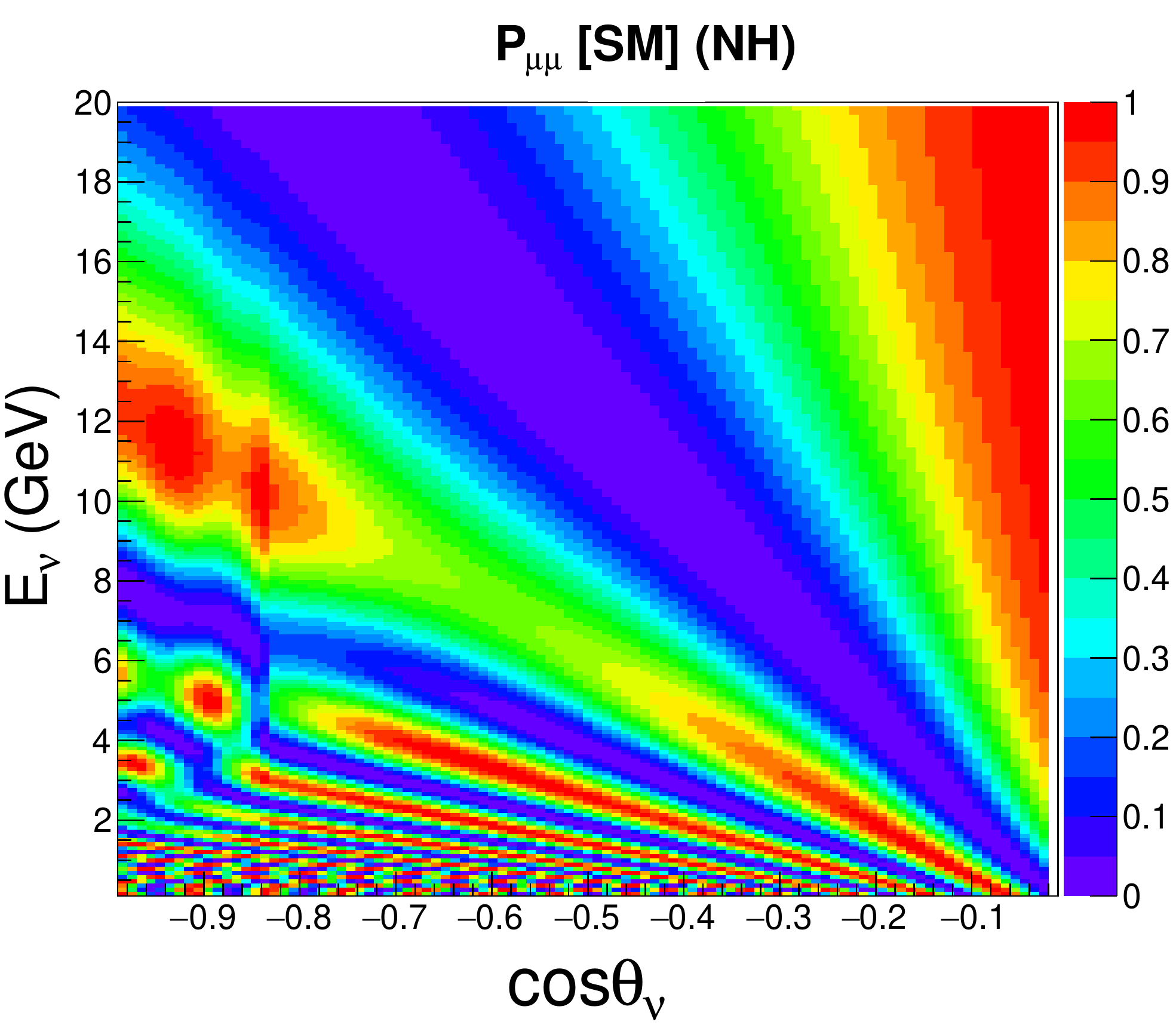}}
 \subfigure{\includegraphics[width=0.32\linewidth]{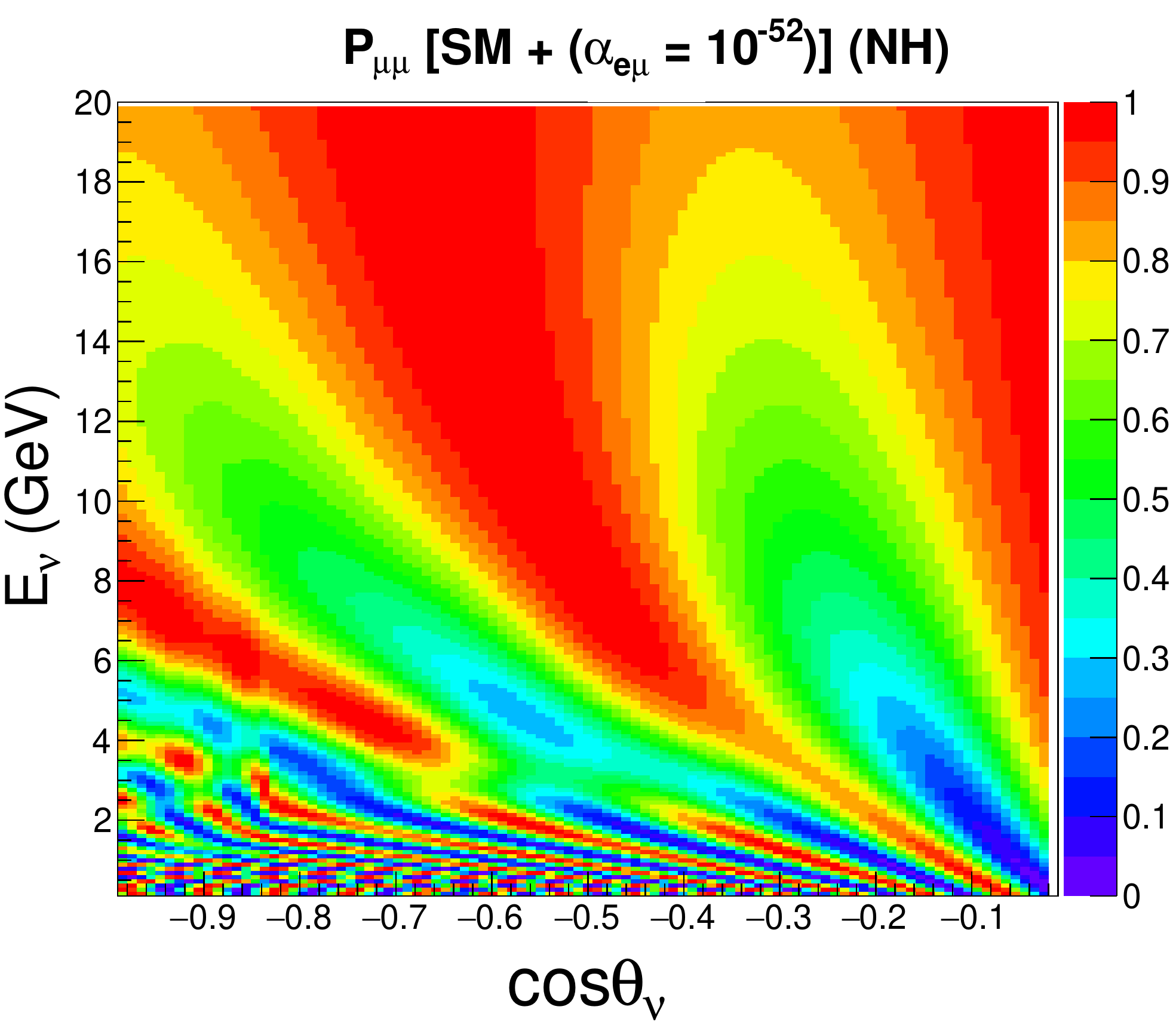}}
 \subfigure{\includegraphics[width=0.32\linewidth]{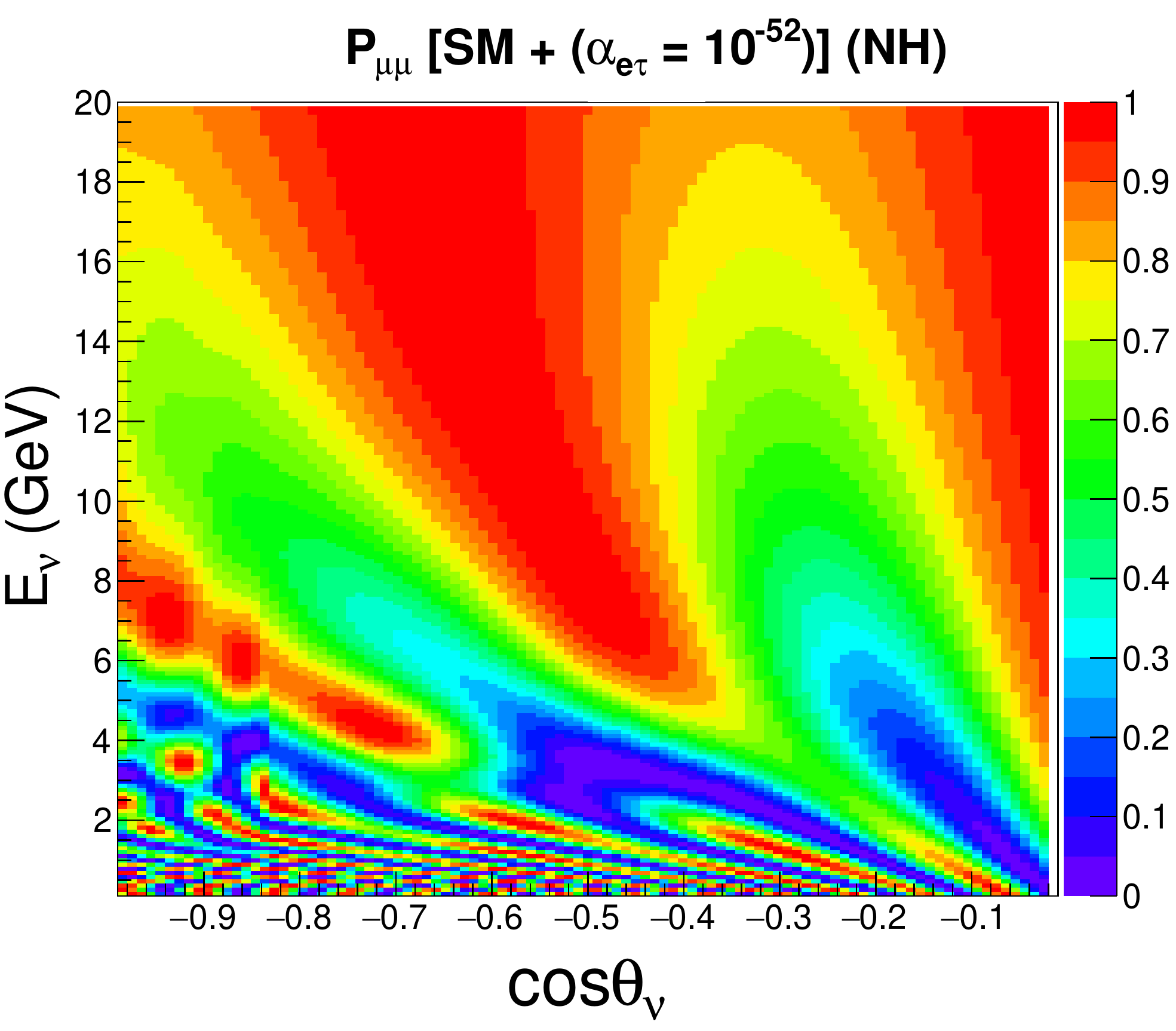}}
  \mycaption{ The oscillograms for $\nu_\mu\rightarrow\nu_\mu$ channel in  $E_\nu$, $\cos\theta_{\nu}$ plane
 for three different scenario: i) $\alpha_{e\mu}=\alpha_{e\tau} =0$ (the SM case, left panel), ii) $\alpha_{e\mu} 
 = 10^{-52}$, $\alpha_{e\tau} =0$ (middle panel), and iii) $\alpha_{e\mu}  = 0$, $\alpha_{e\tau} =10^{-52}$ 
 (right panel). Here, in all the panels, we assume NH. }
 \label{fig:osc-pmumu}
 \end{figure}
In Fig.\,\ref{fig:osc-pmumu}, we present the oscillograms for $\nu_\mu$ survival 
channel in the plane of $\cos\theta_\nu$ vs. $E_\nu$ considering NH. Here, we draw the oscillograms 
for the same three cases as considered in Fig.\,\ref{fig:osc-pemu}. 
First, we notice that for $E_\nu$ in the range of 6 to 20 GeV and $\cos\theta_\nu$ in the range 
of -1 to -0.2, survival probability $P_{\mu\mu}$ gets enhanced significantly for both 
non-zero $\alpha_{e\mu}$ (middle panel) and $\alpha_{e\tau}$ (right panel) as compared to the SM case 
(see left panel). The reason is the following. As we move to higher energies, $\theta^m_{23}$ 
deviates from maximal mixing for both non-zero $\alpha_{e\mu}$ and $\alpha_{e\tau}$. As a result,   
the term $\sin^2 2\theta^m_{23}$ in Eq.\,\ref{eq:pmumu} gets reduced and causes an enhancement 
in $P_{\mu\mu}$.  In Fig.\,\ref{fig:osc-pmumu}, we see some differences in the oscillogram 
patterns in the energy range of 2 to 5 GeV for $L_e-L_\mu$ (middle panel) and $L_e-L_\tau$ 
(right panel) symmetries. Let us try to understand the reason behind this. We have already seen 
that $\theta^m_{23}$ ``runs'' in the opposite directions from $45^\circ$ for $L_e-L_\mu$ 
and $L_e-L_\tau$ symmetries. Due to this, the only term $\frac{1}{4}\tan^2\theta^m_{23}\,\sin^2 2\theta^m_{13}$ 
in Eq.\,\ref{eq:pmumu} gives different contributions for finite $\alpha_{e\mu}$ and $\alpha_{e\tau}$. 
Around the resonance region ($E\sim$ 2 to 5 GeV), $\theta^m_{13}$ attains the maximal value, and the 
strength of above mentioned term becomes quite significant which causes the differences in $P_{\mu\mu}$ 
for these two U(1) symmetries under consideration. We see the effect of this feature in the top 
left panel of Fig.\,\ref{fig:event-letau-lemu}, which we discuss later.

\section{Important Features of the ICAL detector}
\label{sec:ical}
The proposed Iron Calorimeter (ICAL) detector\,\cite{Kumar:2017sdq} under the  
India-based Neutrino Observatory (INO)\,\cite{INO} project plans to study the 
fundamental properties of atmospheric neutrino and antineutrino separately using 
the magnetic field inside the detector. The strength of the magnetic field will 
be around 1.5 T  with a better uniformity in the central region\,\cite{Behera:2014zca}.
It helps to determine the charges of $\mu^-$ and $\mu^+$ particles which get produced 
in the charged-current (CC) interactions of $\nu_\mu$ and $\bar\nu_\mu$ inside the 
ICAL detector. To restrict the cosmic muons, which serve as background in our case, 
the ICAL detector is planned to have rock coverage of more than 1 km all around.
According to the latest design of the ICAL detector\,\cite{Kumar:2017sdq,INO}, 
it consists of alternate layers of iron plates and glass Resistive Plate Chambers 
(RPCs)\,\cite{Datar:2009zz}, which act as the target material and active detector  
elements respectively. While passing through the RPCs, the minimum ionizing particle 
muon gives rise to a distinct track, whose path is recorded in terms of strip hits.  
We identify these tracks with the help of a track finding algorithm.  
Then, we reconstruct the momentum and charge of muon using the well known  
Kalman Filter\,\cite{Bhattacharya:2014tha,Bhattacharya:2015odk} package.
The typical detection efficiency of a 5 GeV muon in ICAL traveling vertically 
is around $80 \%$, while the achievable charge identification efficiency is more 
than $95 \%$\,\cite{Chatterjee:2014vta}. In ICAL, the energy resolution ($\sigma/E$) of a 
5 to 10 GeV muon varies in the range of $10\%$ to $15\%$, while its direction may be 
reconstructed with an accuracy of one degree\,\cite{Chatterjee:2014vta}. 
The prospects of ICAL to measure the three-flavor oscillation parameters based on 
the observable ($E_\mu,\cos\theta_\mu$) have already been studied in Ref.\,\cite{Ghosh:2012px,
Thakore:2013xqa}. 

The hits in the RPCs due to hadrons produce shower-like features. Recently, the possibilities of 
detecting hadron shower and measuring its energy in  ICAL have been explored\,\cite{Devi:2013wxa,
Mohan:2014qua}. These final state hadrons get produced along with the muons in CC deep-inelastic 
scattering process in multi-GeV energies, and can provide vital information about 
the initial neutrino. We can calibrate the energy of hadron ($E^{'}_{had} = E_\nu - E_\mu$) 
using number of hits produced by hadron showers\,\cite{Devi:2013wxa}. Preliminary studies have 
shown that one can achieve an energy resolution of 85$\%$ ($36\%$) at 1 GeV (15 GeV). 
Combining the muon ($E_\mu,\cos\theta_\mu$) and hadron ($E^{'}_{had} $) information on an 
event-by-event basis, the physics reach of ICAL to the neutrino oscillation parameters 
can be improved significantly\,\cite{Devi:2014yaa}. We follow the Refs.\,\cite{Chatterjee:2014vta}  
and \cite{Devi:2013wxa} to incorporate the detector response for muons  and hadrons respectively.

\section{Event Spectrum in the ICAL Detector} 
\label{sec:event}
  \begin{figure}[htb!]
  \subfigure{\includegraphics[width=0.325\linewidth]{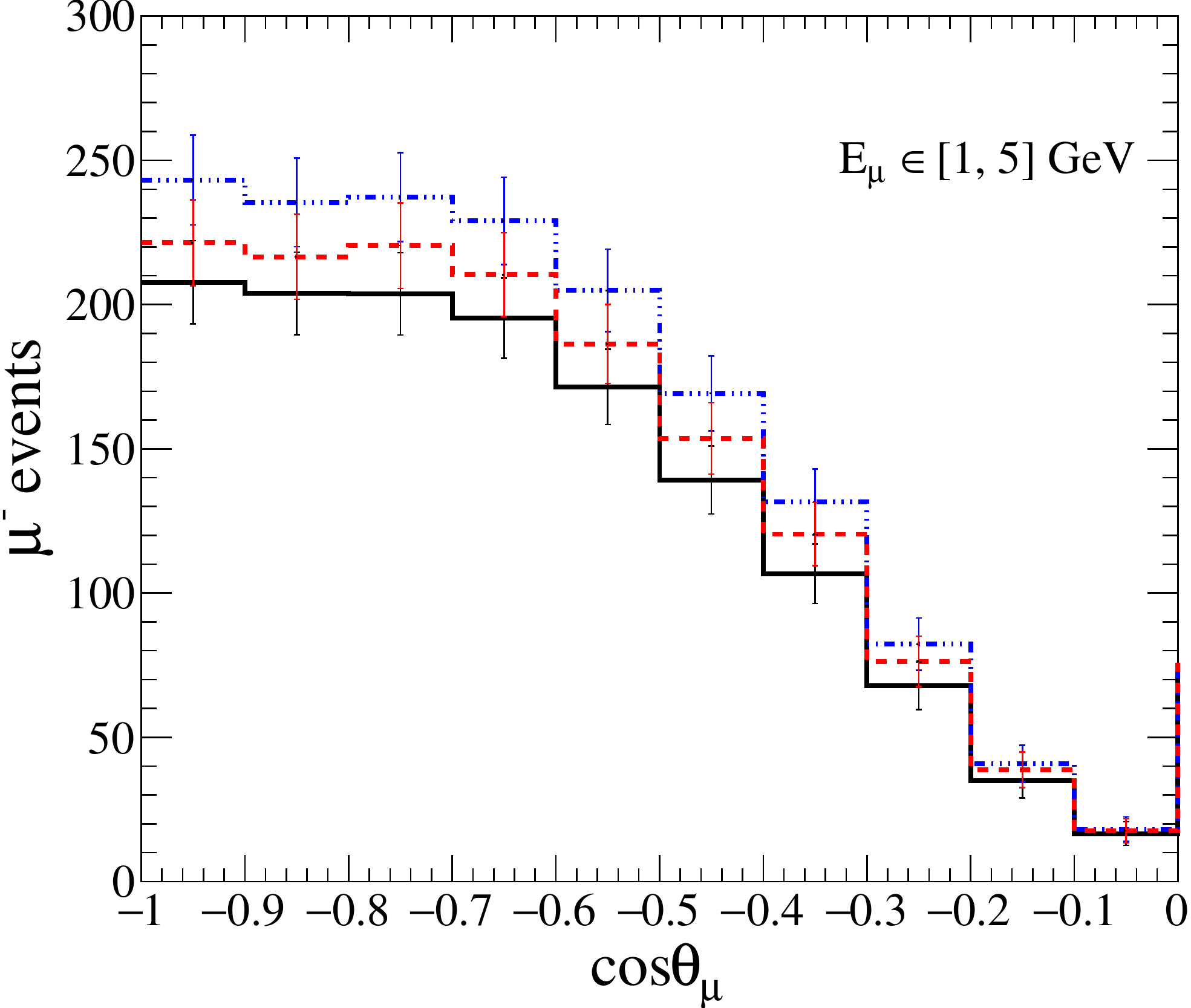}}
  \subfigure{\includegraphics[width=0.325\linewidth]{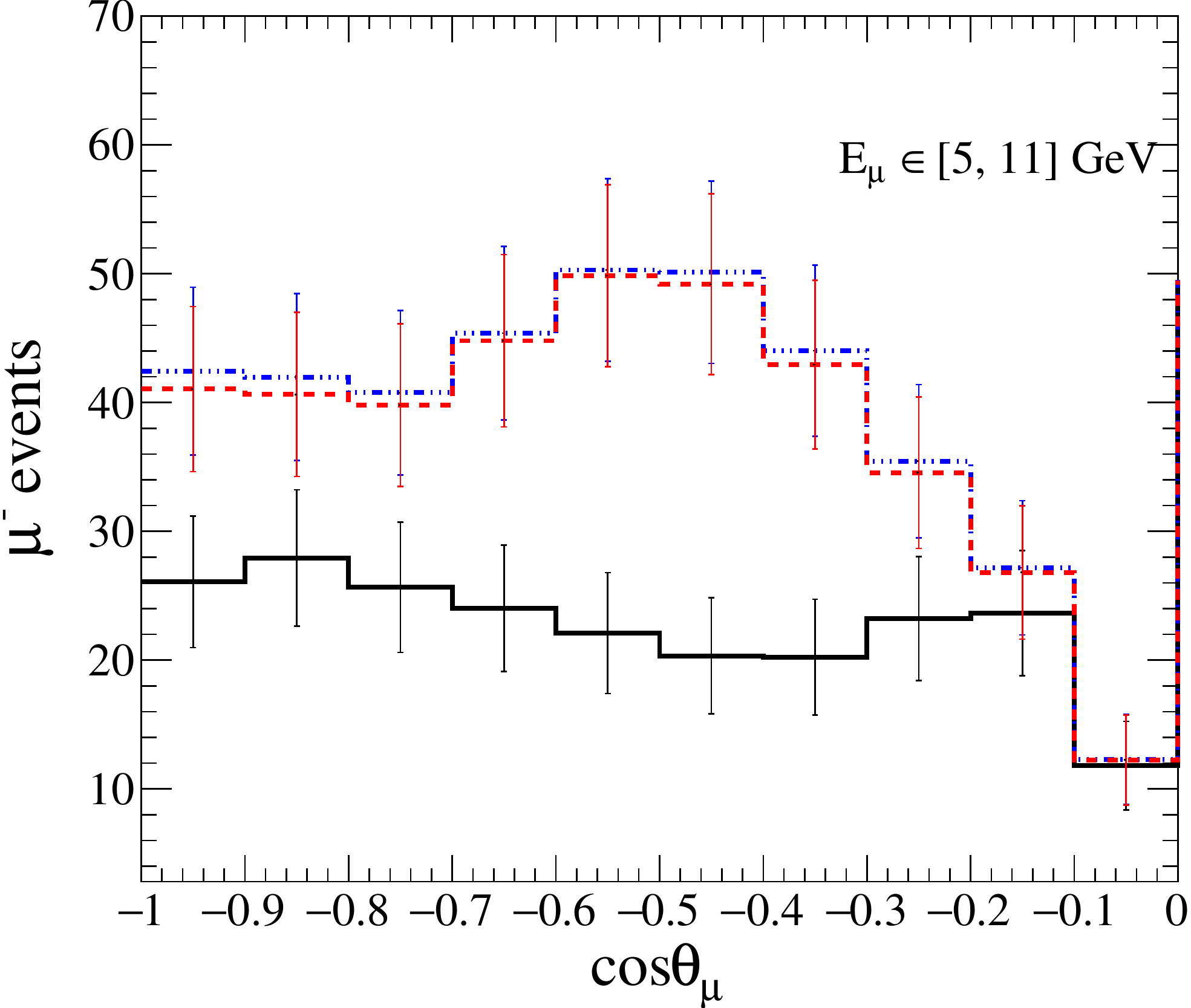}}
  \subfigure{\includegraphics[width=0.325\linewidth]{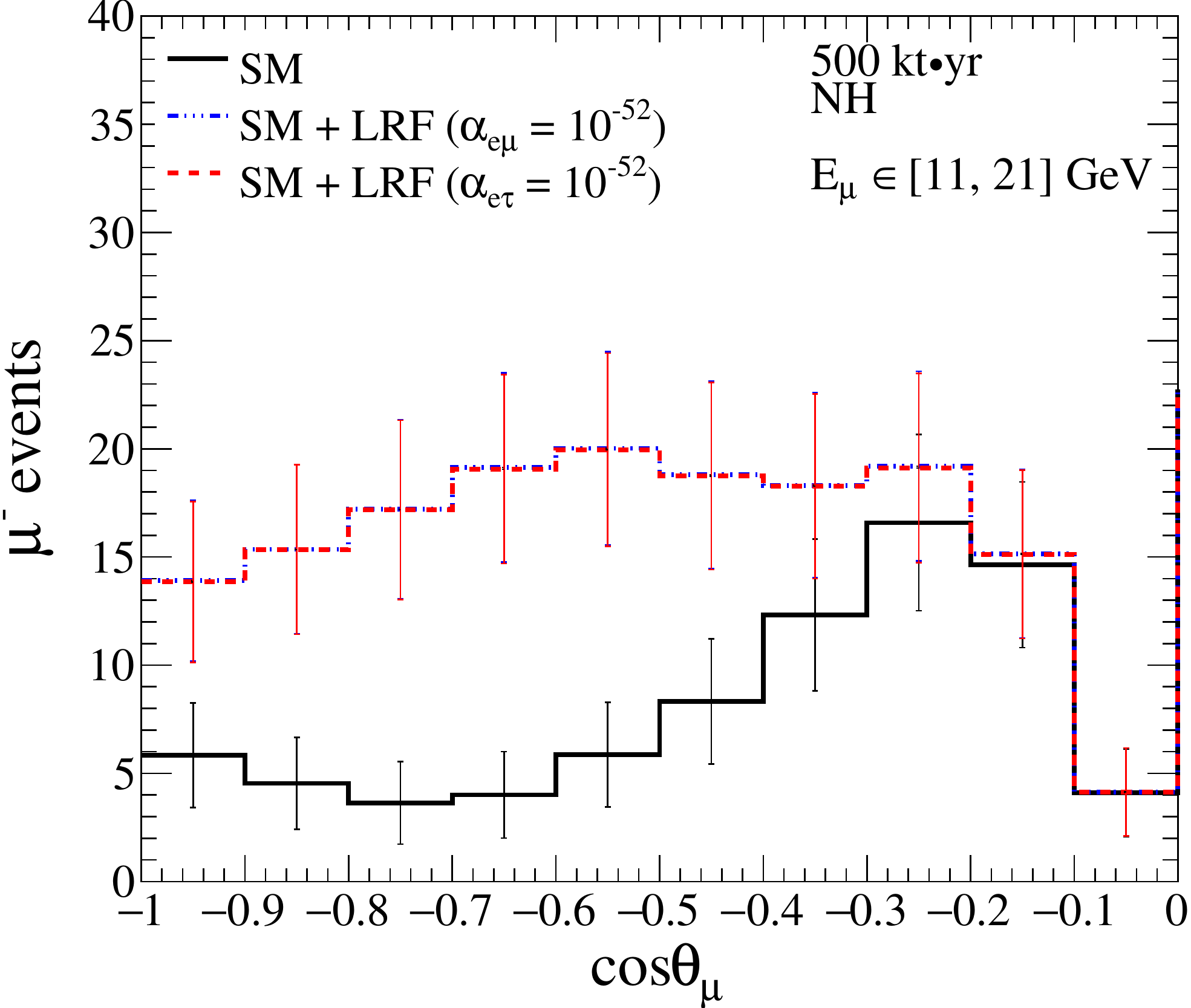}}
  \subfigure{\includegraphics[width=0.325\linewidth]{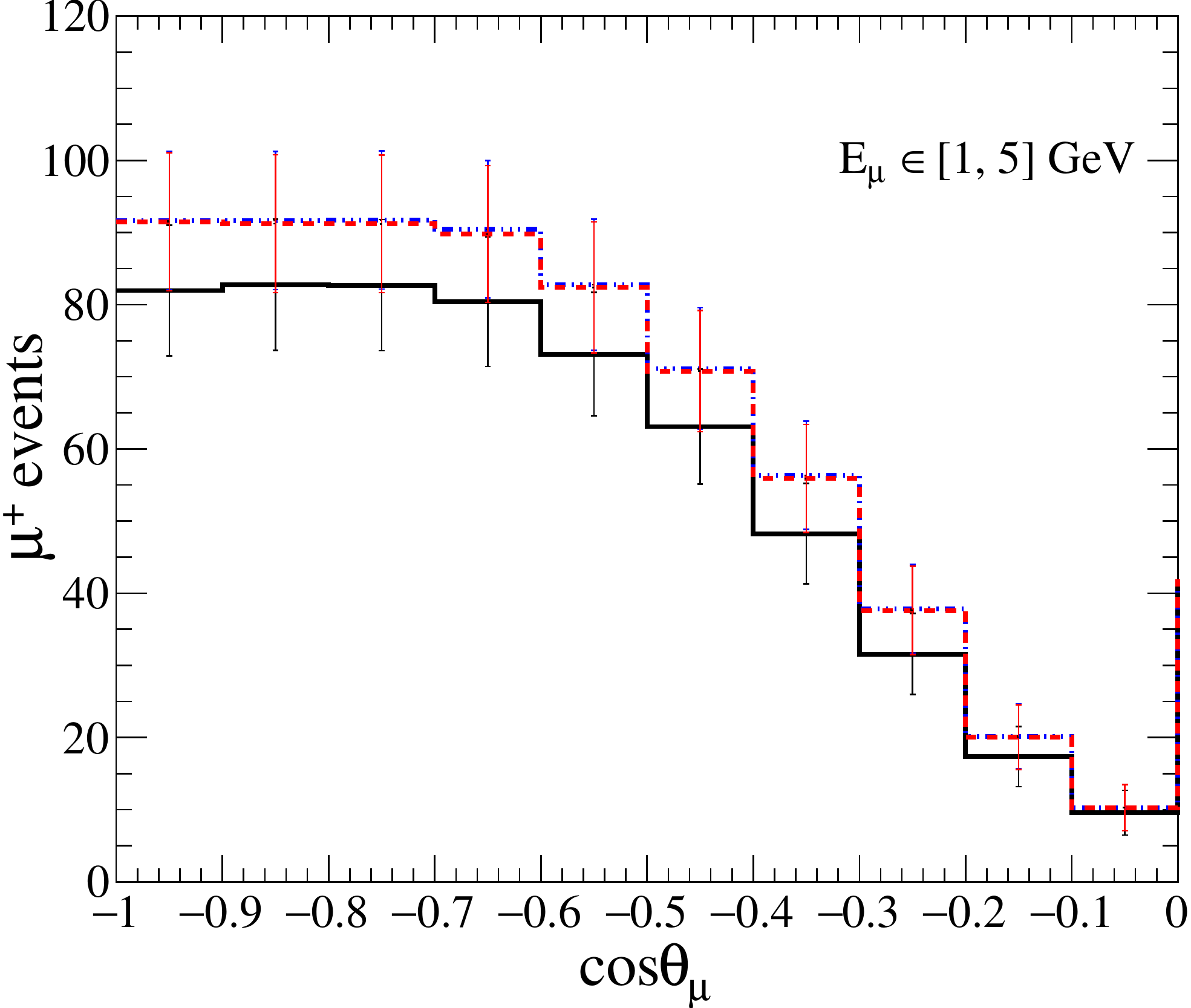}}
  \subfigure{\includegraphics[width=0.325\linewidth]{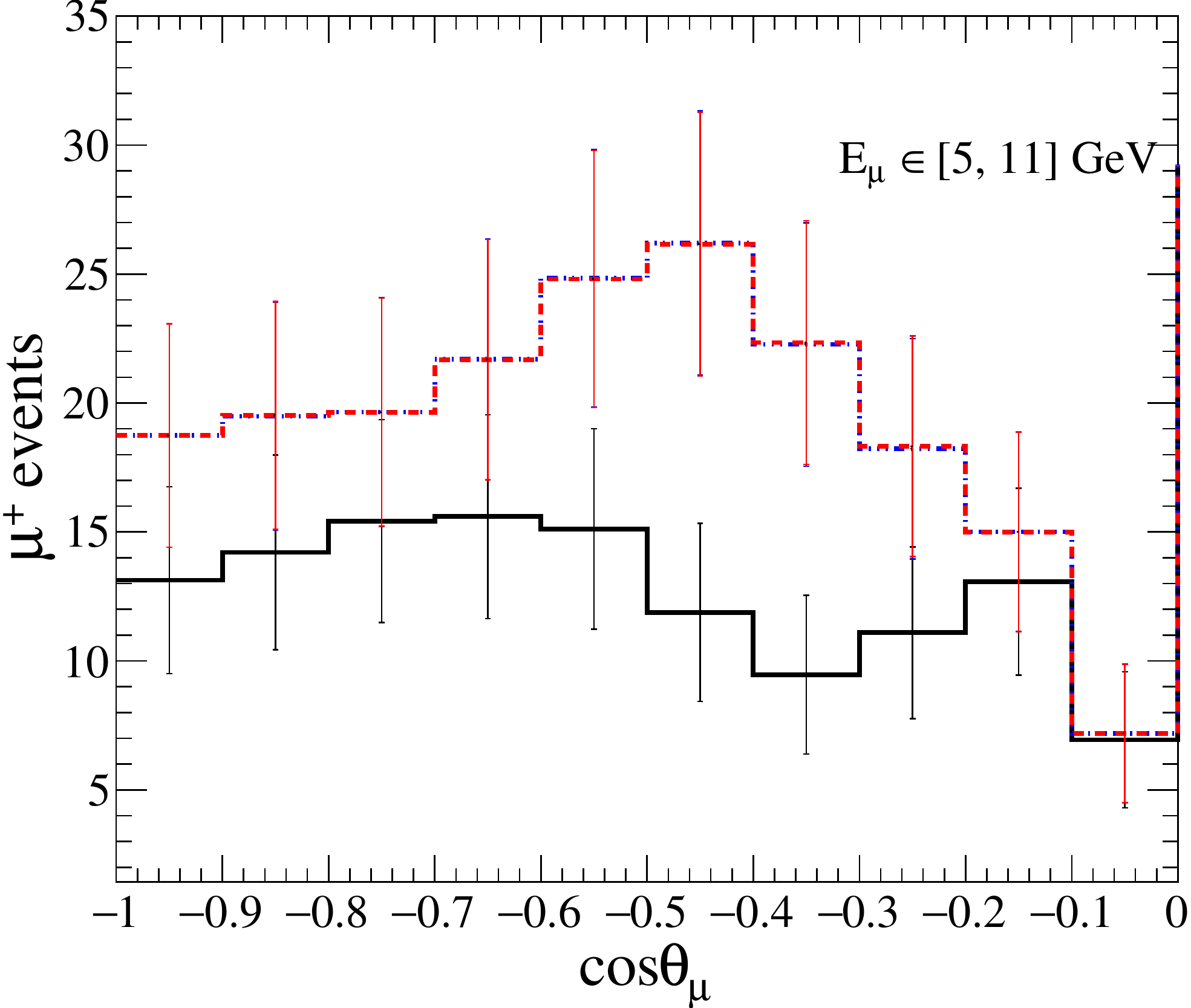}}
  \subfigure{\includegraphics[width=0.325\linewidth]{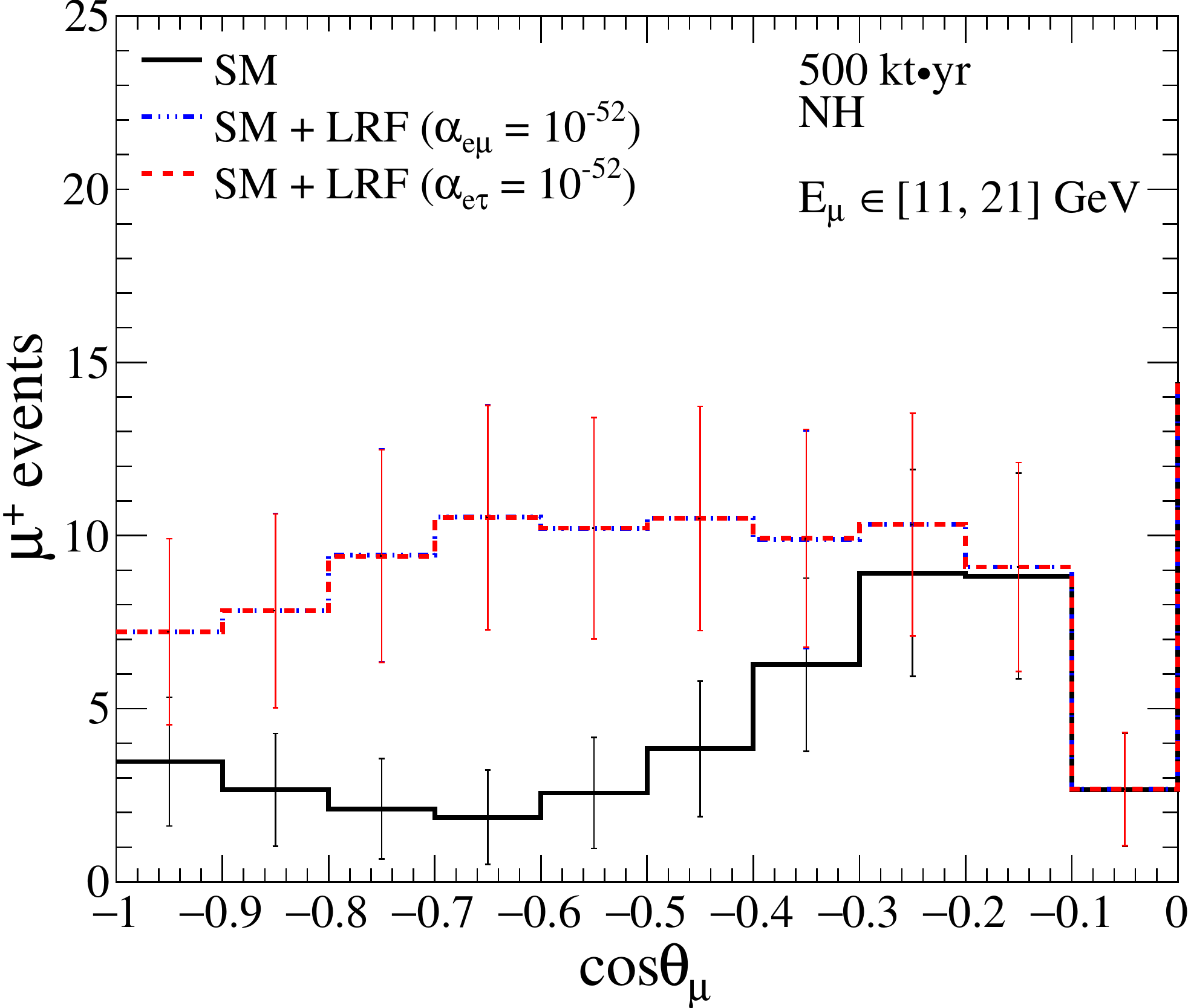}}
  \mycaption{The distributions of $\mu^-$ (upper panels) and $\mu^+$ (lower panels) events 
  for three different $E_\mu$ bins: 1 to 5 GeV in left panel, 5 to 11 GeV in middle panel, and 
   11 to 21 GeV in right panel. 
  In each panel, we consider three different cases: i) $\alpha_{e\mu} = \alpha_{e\tau} =0$ (the SM 
  case, black solid line), ii)  $\alpha_{e\mu} = 10^{-52}$, $\alpha_{e\tau} =0$ (blue dash-dotted line), and 
iii) $\alpha_{e\mu} = 0$, $\alpha_{e\tau} =10^{-52}$ (red dashed line). Here, we sum over $E^{'}_{had}$ in its 
entire range of 0 to 25 GeV and show the results for 500 kt$\cdot$yr exposure and assuming NH. }
  \label{fig:event-letau-lemu}
  \end{figure}
In this section, we present the expected event spectra and total event rates in ICAL with 
and without long-range forces. Using the event generator NUANCE\,\cite{Casper:2002sd} 
and atmospheric neutrino fluxes at Kamioka\footnote{Preliminary calculation of the expected 
fluxes at the INO site have been performed in Ref.\,\cite{Athar:2012it}. We plan to use these 
fluxes in future analysis once they are finalized. The horizontal components of the geo-magnetic 
field are different at the INO (40 $\mu$T) and Kamioka (30 $\mu$T). Due to this reason, 
we observe a difference in atmospheric fluxes at these sites.} \cite{Honda:2011nf}, we obtain 
the unoscillated event spectra for neutrino and antineutrino. After incorporating the detector 
response for muons and hadrons as described in Ref.\,\cite{Devi:2014yaa} and for the benchmark 
values of the oscillation parameters as mentioned in section\,\ref{subsec:running-exp} 
($\sin^2\theta_{23}=0.5$, $\,\sin^2 2\theta_{13}= 0.0847$, and NH), we obtain around 4870 (2187) $\mu^-$ 
($\mu^+$) events for the SM case using a 500 kt$\cdot$yr exposure. To obtain these event rates, we consider   
$E_\mu$ in the range 1 to 21 GeV, $\cos\theta_\mu$ in its entire range of -1 to 1, 
and $E^{'}_{had}$ in the range 0 to 25 GeV. In presence of $L_e-L_\mu$ symmetry with $\alpha_{e\mu}=10^{-52}$, 
the number of $\mu^-$ ($\mu^+$) events becomes 5365 (2373). For $L_e-L_\tau$ symmetry with 
$\alpha_{e\tau}=10^{-52}$, we get 5225  $\mu^-$ and 2369 $\mu^+$ events. In Fig.\,\ref{fig:event-letau-lemu},
we show the distribution of only upward going $\mu^-$ (top panels) and $\mu^+$ (bottom panels) 
events as a function of reconstructed  $\cos\theta_\mu$ in the range -1 to 0. Here, we integrate 
over the entire range of hadron energy ($E^{'}_{had} \in$ 0 to 25 GeV), and display the event 
spectra considering three different $E_\mu$ bins having the ranges 1 to 5 GeV (left panels), 5 to 11 GeV 
(middle panels), and 11 to 21 GeV (right panels). In each panel, we compare the event distribution 
for three different scenarios: i) $\alpha_{e\mu}=\alpha_{e\tau} =0$ (the SM case, black solid lines), 
ii) $\alpha_{e\mu}  = 10^{-52}$, $\alpha_{e\tau} =0$ (blue dash-dotted lines), and iii) $\alpha_{e\mu}  
= 0$, $\alpha_{e\tau} =10^{-52}$ (red dashed lines). 
We observe few interesting features in Fig.\,\ref{fig:event-letau-lemu}, which we discuss now. 

In all the panels of Fig.\,\ref{fig:event-letau-lemu}, we see an enhancement in the event rates 
for $\cos\theta_\mu\in$ -1 to -0.2 in the presence of long-range forces as compared to the SM case. 
This mainly happens due to substantial increase in $P_{\mu\mu}$ with finite $\alpha_{e\mu}$ 
or $\alpha_{e\tau}$ as compared to the SM case. We have already seen this feature in Fig.\,\ref{fig:osc-pmumu}. 
Also, we see similar event distributions for both the symmetries in all the panels, except 
in the top left panel ($E_\mu\in$ 1 to 5 GeV), where we see some differences in the event spectra for 
$L_e-L_\mu$ and $L_e-L_\tau$ symmetries. We have already explained the reason behind this with the help of 
oscillogram patterns (see middle and right panels in Fig.\,\ref{fig:osc-pmumu}) in section\,\ref{subsec:osc-pmumu}.
Next, we discuss the binning scheme for three observables ($E_\mu$, $\cos\theta_\mu$, and $E'_{had}$), and briefly 
describe the numerical technique and analysis procedure which we 
adopt to estimate the physics reach of ICAL.  

\section{Simulation Procedure}
\label{sec:sim-proc}
\subsection{Binning Scheme for Observables ($E_\mu$, $\cos\theta_\mu$, $E'_{had}$)}
\label{subsec:bin}
 \begin{table}[htb!] 
\centering 
\begin{tabular}{|c| c| c| c| c|} 
\hline\hline 
Observable  & Range & Bin width & No. of bins & Total bins \\ 
\hline
$E_{\mu}$ (GeV) & \makecell[c]{ $[1,11]$ \\$[11,21]$} & 
\makecell[c]{ 1 \\ 5} & 
\makecell{ 10 \\2 } & \makecell{ 12 } 
 \\
\hline
$\cos\theta_\mu$  & \makecell[c]{ $[-1.0,0.0]$ \\ $[0.0,1.0]$ } 
& \makecell[c]{ 0.1 \\ 0.2} & 
\makecell[c]{10 \\ 5} & \makecell[c]{15 }\\
\hline
$E'_{\rm had}$ (GeV)  
& \makecell[c]{$[0,2]$ \\ $[2,4]$ \\  $[4,25]$} 
& \makecell[c]{1 \\ 2 \\21 } & 
\makecell[c]{2 \\ 1 \\1} & \makecell[c]{ 4 }\\
\hline\hline
\end{tabular}
\mycaption{The binning scheme considered for the reconstructed observables  
$E_\mu$, $\cos\theta_\mu$, and $E'_{had}$ for each muon polarity. In last column,
we give the total number of bins taken for each observable.}
\label{tab:bin}
\end{table}
Table\,\ref{tab:bin} shows the binning scheme that we adopt in our simulation for three observables 
$E_\mu$ ($\in$ 1 to 21 GeV), $\cos\theta_\mu$ ($\in$ -1 to 1), and  $E'_{had}$ ($\in$ 0 to 25 GeV). 
In these ranges, we have total 12 bins for $E_\mu$, 15 bins for $\cos\theta_\mu$, and 4 bins for 
$E'_{had}$, resulting into a total of ($12\times15\times4=$) 720 bins per polarity.  We consider the 
same binning scheme for $\mu^-$ and $\mu^+$ events. As we go to higher energies, the atmospheric neutrino 
flux decreases resulting in lower statistics. Therefore, we take wider bins for $E_\mu$ and $E'_{had}$
at higher energies. We do not perform any optimization study for binning, however we make sure 
that we have sufficient statistics in most of the bins without diluting the sensitivity much. 
In our study, the upward going events ($\cos\theta_\mu$ in the range 0 to -1) play an 
important role, where $V_{CC}$, $V_{e\mu/e\tau}$, and $\Delta m^2_{31} /2E$ become 
comparable and can interfere with each other (see discussion in section\,\ref{sec:framework}). 
Therefore, we take 10 bins of equal width for upward going events which is compatible 
with the angular resolutions of muon achievable in ICAL. The downward going events do not 
undergo oscillations. But, they certainly enhance the overall statistics and help us to
reduce the impact of normalization uncertainties in the atmospheric neutrino fluxes. 
Therefore, we include the downward going events in our simulation considering five  
$\cos\theta_\mu$ bins of equal width in the range of 0 to 1.  

\subsection{Numerical Analysis}
\label{subsec:num-analysis}
In our numerical analysis, we suppress the statistical fluctuations of the ``observed'' event 
distribution. We generate\footnote{For further details regarding the event generation and 
inclusion of oscillation, see Refs.\,\cite{Ghosh:2012px,Thakore:2013xqa,Devi:2014yaa}. } events 
using NUANCE for an exposure of 50000 kt$\cdot$yr. Then, we implement the detector response and 
finally, normalize the event distribution to the actual exposure. This method along with the 
$\chi^2$ function gives us the median sensitivity of the experiment in the frequentist 
approach\,\cite{Blennow:2013oma}.  We use the following  Poissonian $\chi^2_{-}$ for $\mu^-$ events 
in our statistical analysis: 
\begin{equation}
 \chi^2_{-}\,=\,\min_{\zeta_l}\,\sum^{N_{E'_{had}}}_{i=1} 
 \sum^{N_{E_\mu}}_{j=1} \sum^{N_{{\cos\theta_\mu}}}_{k=1}
 \,2\,\bigg[ N^{\rm theory}_{ijk}\,-
 \, N^{\rm data}_{ijk}\, - \,  N^{\rm data}_{ijk}\,\,
 {\rm{ln}}\bigg(\frac{N^{\rm theory}_{ijk}}{N^{\rm data}_{ijk}}
 \bigg)\bigg]\,+\,\sum^5_{l=1} \zeta^2_{l}\,,
\label{eq:chisq}
\end{equation}
with 
\begin{equation}
 N^{\rm theory}_{ijk}\,=\,N^0_{ijk}\big(\,1\,+\,\sum^5_{l=1}
 \pi^l_{ijk}\zeta_{l}\,\big)\,.
\end{equation}
In the above equation, $N^{\rm data}_{ijk}$ and $N^{\rm theory}_{ijk}$ denote the ``observed'' and 
expected number of $\mu^-$ events in a given ($E_\mu$, $\cos\theta_\mu$, $E'_{had}$) bin. $N^0_{ijk}$ 
represents the number of events without systematic uncertainties. In our simulation, $N_{E'_{had}}=4$, 
$N_{E_\mu}=12$, and $N_{{\cos\theta_\mu}}=15$ (see table\,\ref{tab:bin}). We obtain $ N^{\rm data}_{ijk}$ 
using the benchmark values of the oscillation parameters as mentioned in section\,\ref{subsec:running-exp}
and assuming normal hierarchy as neutrino mass hierarchy.
We consider five systematic errors in our analysis: 20$\%$ flux normalization error, 10$\%$ error in cross-section, 
5$\%$ tilt error, 5$\%$ zenith angle error, and 5$\%$ overall systematics. We incorporate these systematic 
uncertainties in our simulation using the well known ``pull'' method\,\cite{Huber:2002mx,Fogli:2002pt,
GonzalezGarcia:2004wg}. 

In a similar fashion, we obtain $\chi^2_+$ for $\mu^+$ events. We estimate the total $\chi^2$ by adding 
the individual contributions coming from $\mu^-$ and $\mu^+$ events in the following way 
\begin{equation}
\chi^2_{\rm ICAL} = \chi^2_-\,+\,\chi^2_+\,.
\label{eq:total-chisq}
\end{equation}
In the fit, we first minimize $\chi^2_{\rm ICAL}$ with respect to the pull variables $\zeta_l$,  
and then marginalize over the oscillation parameters $\sin^2\theta_{23}$ in the range $0.38$ 
to  $0.63$ and $\Delta m^2_{31}$ in the range 0.0024 eV$^2$ to 0.0026 eV$^2$. While deriving the 
constraints on $\alpha_{e\mu/e\tau}$, we also marginalize $\chi^2_{\rm ICAL}$ over both NH and IH.  
We do not marginalize over $\Delta m^2_{21}$, $\sin^2\theta_{12}$, and $\sin^2 2 \theta_{13}$ since 
these parameters are already measured with high precision, and the existing uncertainties on these parameters 
do not alter our results. We consider $\delta_{\rm CP} = 0^\circ$ throughout our analysis. 

\begin{figure}[htb!]
 \subfigure{\includegraphics[width=0.49\linewidth]{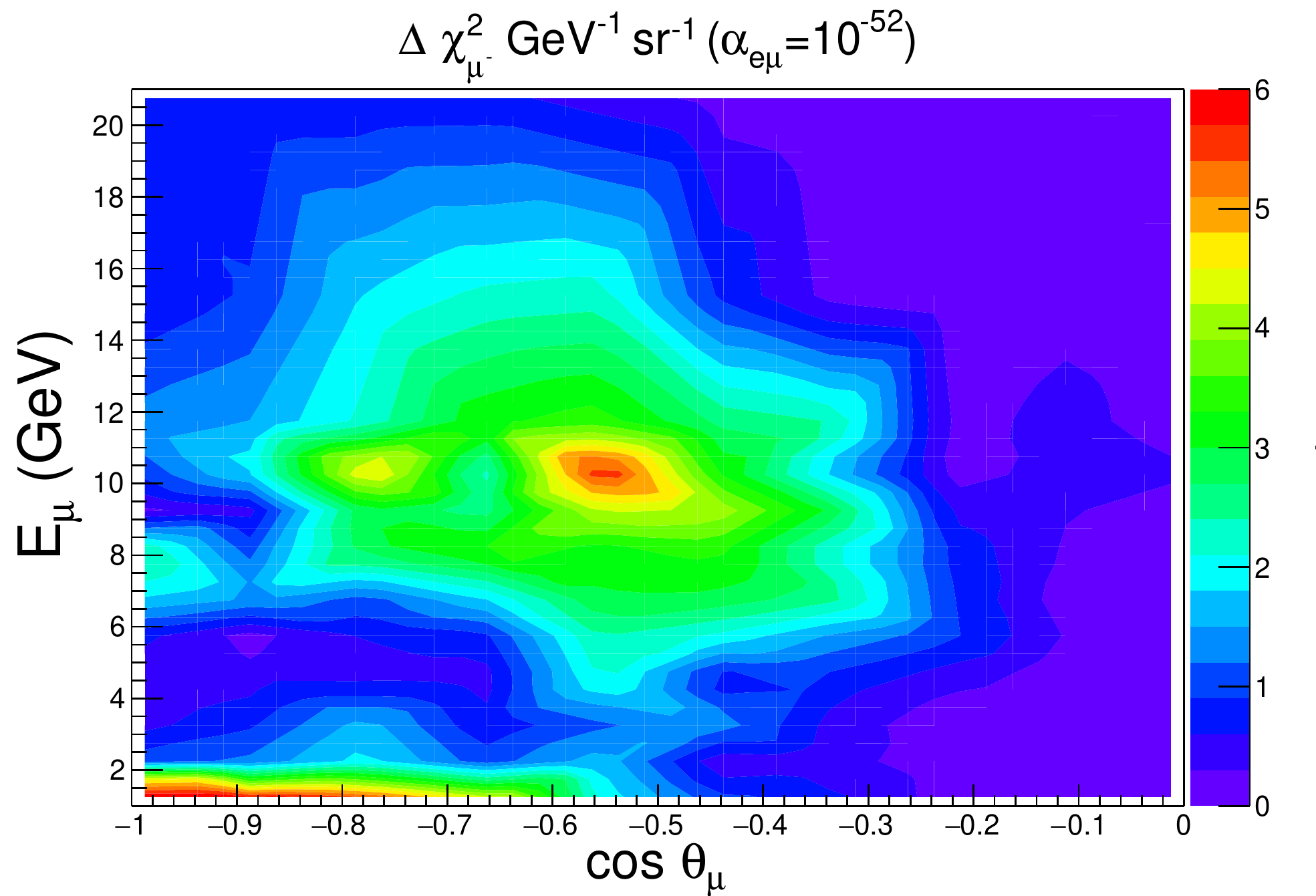}}
 \subfigure{\includegraphics[width=0.49\linewidth]{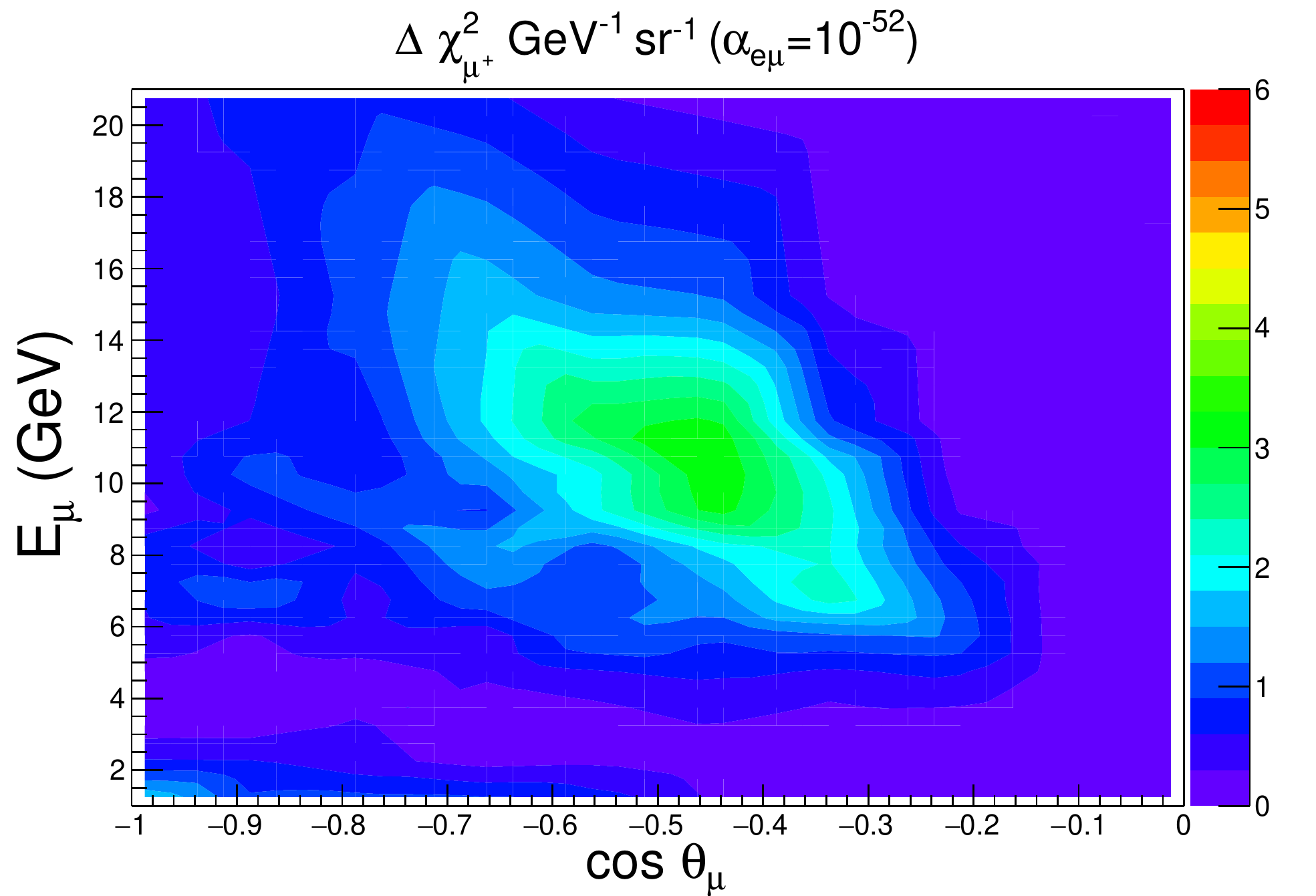}}
  \subfigure{\includegraphics[width=0.49\linewidth]{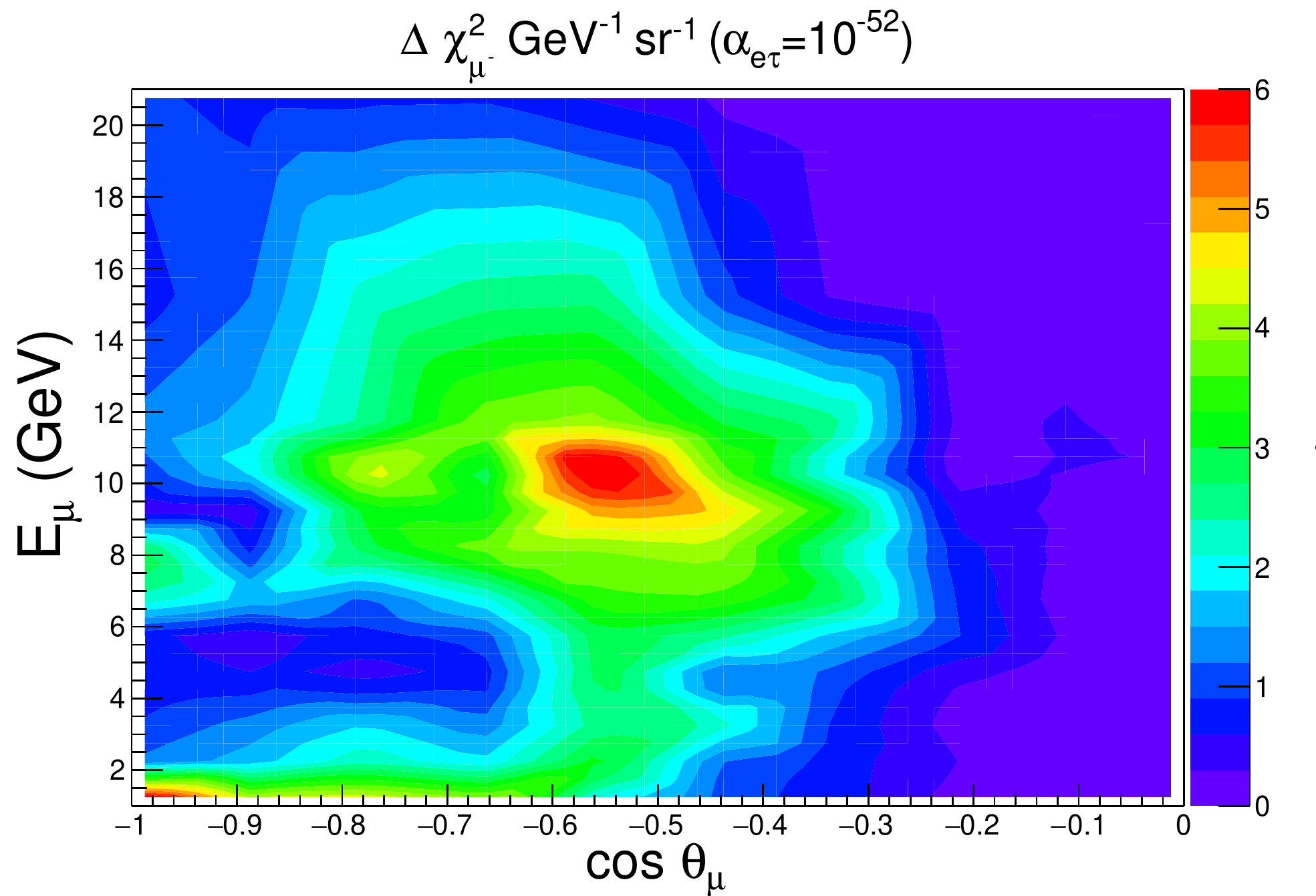}}
 \subfigure{\includegraphics[width=0.49\linewidth]{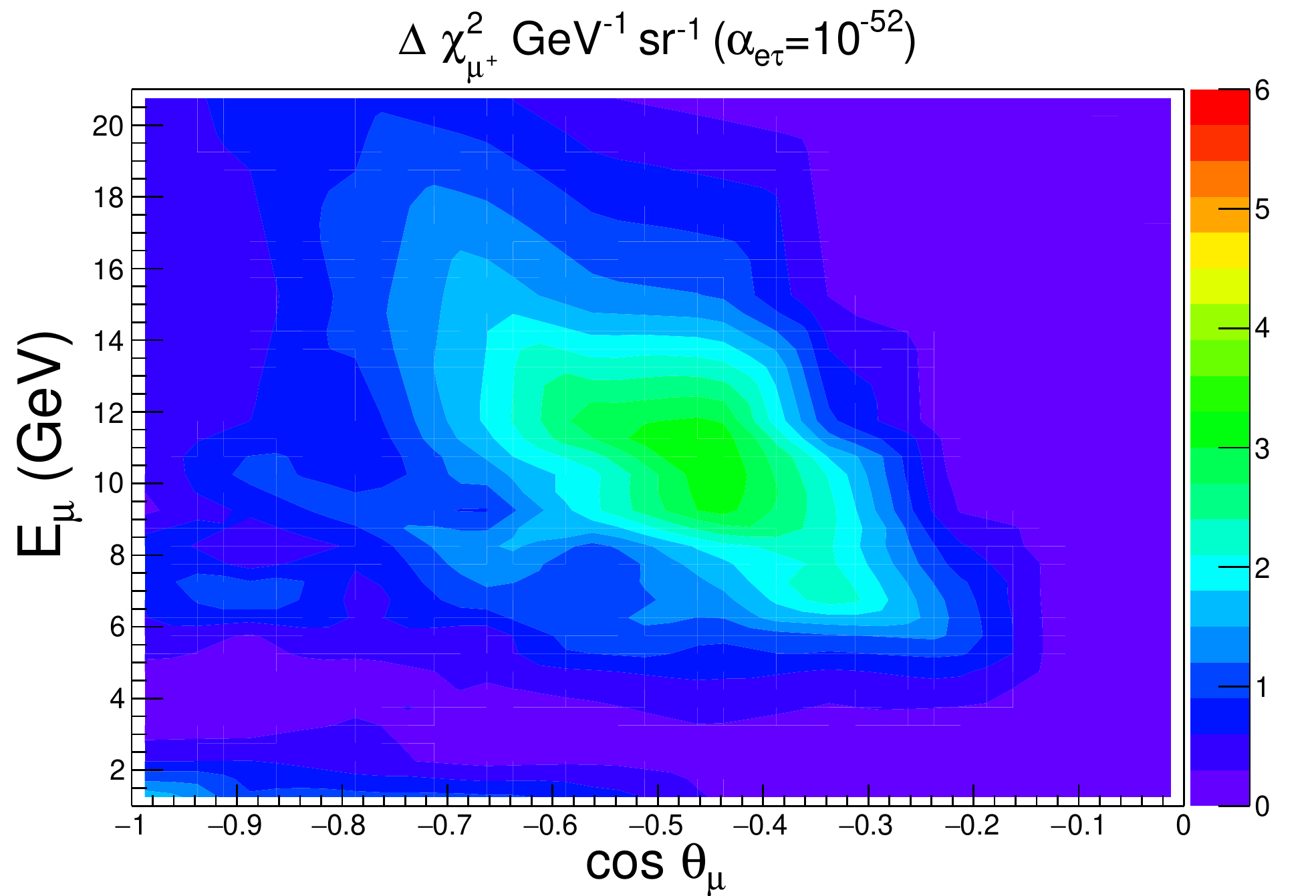}}
 \mycaption{ Distributions of $\Delta\chi^2_{\rm ICAL-LRF}$ (per unit area) in $E_\mu$ and 
 $\cos\theta_\mu$ plane. The left (right) panels are for $\mu^-$ ($\mu^+$) events.  
 In upper (lower) panels, we assume  non-zero $\alpha_{e\mu}$ ($\alpha_{e\tau}$)  
 in the fit with a strength of $10^{-52}$. In all the panels, we use 500 kt$\cdot$yr 
 exposure and assume NH in both data and theory. }
 \label{fig:chisq-dist-e-vs-costhe}
 \end{figure}

\section{Results}
\label{sec:results}

We quantify the statistical significance of the analysis to constrain the LRF parameters 
in the following way
\begin{equation}
 \Delta \chi^2_{\,\,{\rm ICAL-LRF}} = \chi^2_{\,\,{\rm ICAL}}\left({\rm SM} + \alpha_{e\mu/e\tau}\right)
 -\chi^2_{\,\,{\rm ICAL}}\left({\rm SM}\right)\,.  
\end{equation}
Here, $\chi^2_{\,\,{\rm ICAL}} ({\rm SM})$ and $\chi^2_{\,\,{\rm ICAL}}\left({\rm SM} + 
\alpha_{e\mu/e\tau}\right)$ are calculated by fitting the ``observed'' data  
in the absence and presence of LRF parameters respectively. 
In our analysis, statistical fluctuations are suppressed, and therefore, $\chi^2_{\,\,{\rm ICAL}}
({\rm SM})\approx 0$. 
Before we present the constraints on $\alpha_{e\mu/e\tau}$, we identify the regions in 
$E_\mu$ and $\cos\theta_\mu$ plane which give significant contributions toward  
$\Delta \chi^2_{\,\,{\rm ICAL-LRF}}$. In Fig.\,\ref{fig:chisq-dist-e-vs-costhe}, we show 
the distribution\footnote{In Fig.\,\ref{fig:chisq-dist-e-vs-costhe}, we do not consider 
the constant contributions in $\chi^2$ coming from the term 
which involves five pull parameters $\zeta^2_l$ in Eq.\,\ref{eq:chisq}. Also, we do not marginalize over the 
oscillation parameters in the fit to produce these figures. But, we show our final results   
considering full pull contributions and marginalizing over the oscillation parameters in the 
fit as mentioned in previous section.} of $\Delta \chi^2_{\mu^-}$ (left panels) and $\Delta \chi^2_{\mu^+}$ 
(right panels) in the reconstructed $E_\mu$ and $\cos\theta_\mu$ plane, 
where the events are further divided into four sub-bins depending on the reconstructed 
hadron energy (see table\,\ref{tab:bin}). In the upper (lower) panels of Fig.\,\ref{fig:chisq-dist-e-vs-costhe},
we take non-zero $\alpha_{e\mu}$ ($\alpha_{e\tau}$) in the fit with a strength of $10^{-52}$. 
We clearly see from the left 
panels that for $\mu^-$ events, most of the contributions ($\sim 70\%$) stem from the range 
6 to 15 GeV for $E_\mu$ and for $\cos\theta_\mu$, the effective range is -0.8 to -0.4. We see 
similar trend for both the symmetries (see upper and lower panels) and for $\mu^+$ events
(see right panels) as well.

 \begin{figure}[htb!]
 \begin{center}
  \includegraphics[width=0.7\linewidth]{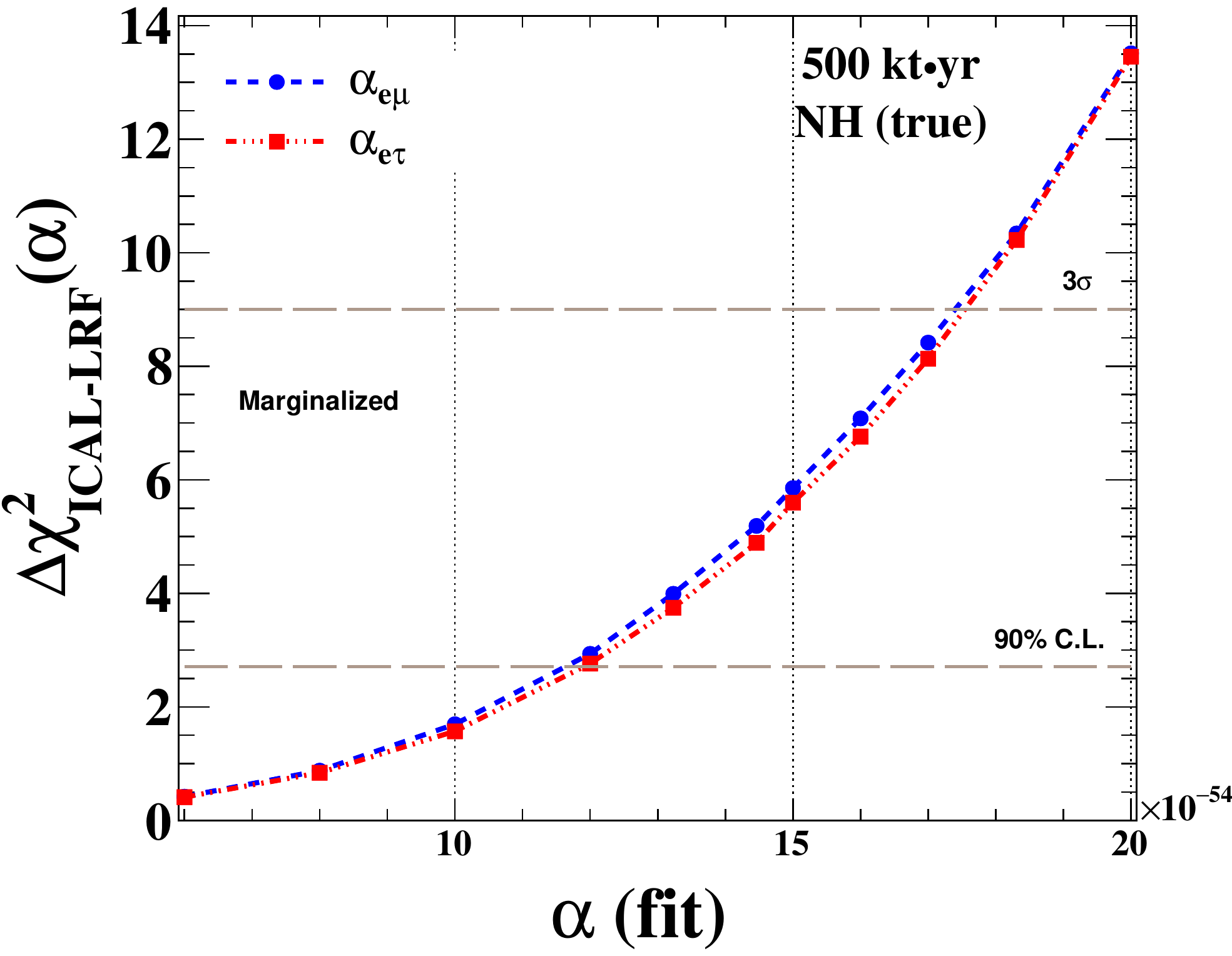}
  \mycaption{Sensitivity of ICAL to set upper limits on $\alpha_{e\mu}$ (blue dashed line) and 
  $\alpha_{e\tau}$ (red dash-dotted line) using 500 kt$\cdot$yr exposure and assuming NH as true choice.}
  \label{fig:lemu-letau-nh}
  \end{center}
 \end{figure}

Fig.\,\ref{fig:lemu-letau-nh} shows the upper bound on $\alpha_{e\mu}$ and $\alpha_{e\tau}$ 
(one at-a-time) using 500 kt$\cdot$yr exposure of ICAL if there is no signal of long-range 
forces in the data. We set new upper limit on $\alpha_{e\mu}$ or $\alpha_{e\tau}$ by 
generating the data with no long-range forces and fitting it with 
some non-zero value of $\alpha_{e\mu/e\tau}$ by means of $\chi^2$ technique as outlined in 
previous section. The corresponding $\Delta \chi^2_{\,\,{\rm ICAL-LRF}}$ obtained after 
marginalizing over $\sin^2\theta_{23}$, $\Delta m^2_{31}$, hierarchy, and systematics parameters 
in the fit, is plotted in Fig.\,\ref{fig:lemu-letau-nh} as a function of $\alpha_{e\mu/e\tau}$ 
(test). It gives a measure of the sensitivity reach of ICAL to the effective gauge coupling of 
LRF. For both the symmetries, we assume NH as true hierarchy. We obtain similar constraints 
for both the symmetries (one at-a-time) since $\alpha_{e\mu}$ and $\alpha_{e\tau}$ affect both 
$P_{\mu\mu}$  and $P_{e\mu}$ oscillation channels in almost similar fashion over a wide range of energies 
and baselines (see Figs.\,\ref{fig:osc-pemu} and \ref{fig:osc-pmumu}). 
The expected upper limit on $\alpha_{e\mu/e\tau}$ from ICAL is $<1.2 \times 10^{-53}$ 
($1.75\times 10^{-53}$) at 90$\%$ (3$\sigma$) C.L. with 500 kt$\cdot$yr exposure and NH as true 
hierarchy. This future limit on $\alpha_{e\mu}$ from ICAL at 90$\%$ C.L. is $\sim$ 46 times better 
than the existing limit from the Super-Kamiokande experiment\,\cite{Joshipura:2003jh}. 
For  $\alpha_{e\tau}$, the limit is 53 times better at 90$\%$ confidence level. We obtain 
similar constraints assuming IH as true hierarchy.  
We see a marginal improvement in the upper limits if we keep all the oscillation 
parameters fixed in the fit. In this fixed parameter case, the new bound becomes 
$\alpha_{e\mu}<1.63\times 10^{-53}$ at $3\sigma$ confidence level. We study few
interesting issues in this fixed parameter scenario which we discuss now. 
\begin{itemize}
 \item \textbf{Advantage of Spectral Information:} In ICAL, we can bin the atmospheric neutrino/antineutrino 
 events in the observables $E_\mu$, $\cos\theta_\mu$, and $E'_{had}$. It helps us immensely 
 to achieve hierarchy measurement at around $3\sigma$ C.L. with 500 kt$\cdot$yr exposure\,\cite{Devi:2014yaa}. 
 We find that the ability of using the spectral information in ICAL also plays an important 
 role to place tight constraint on LRF parameters. For an example, if we rely only on the total $\mu^-$ and 
 $\mu^+$ event rates, the expected limit from ICAL becomes $\alpha_{e\mu}< 2.2\times 10^{-52}$ at 3$\sigma$ 
 confidence level. This limit is almost  13 times weaker as compared to what we can obtain using 
 the full spectral informations.
 
  \item \textbf{Usefulness of Hadron Energy Information:} In our analysis, we use the hadron energy information 
  ($E'_{had}$) along with the muon momentum ($E_\mu$, $\cos\theta_\mu$). We observe that with a value of  
  $\alpha_{e\mu}=1.63\times 10^{-53}$ in the fit, $\Delta\chi^2_{\rm ICAL-LRF}$ increases from 5.2  to 9  
  when we use $E_\mu$, $\cos\theta_\mu$, and $E'_{had}$ as our observables instead of only $E_\mu$ and $\cos\theta_\mu$. 
  It corresponds to about 73$\%$ improvement in the sensitivity. 
   
  \item \textbf{The Role of Charge Identification Capability:}  We also find that the charge identification capability of ICAL in distinguishing $\mu^-$ and $\mu^+$ 
 events does not play an important role to constrain the LRF parameters unlike the mass hierarchy 
 measurements. Since the long-range forces affect the $\mu^-$ and $\mu^+$ event rates in almost 
 similar fashion as compared to the SM case  (see Fig\,\ref{fig:event-letau-lemu}), it is not 
 crucial to separate these events in our analysis in constraining the LRF parameters.   
 
 \end{itemize}
 
 Before we summarize and draw our conclusions in the next section, we make few comments on how 
 the presence of LRF parameters may affect the mass hierarchy measurement in ICAL.
 To perform this study, we generate the data with a given hierarchy and assuming 
 $\alpha_{e\mu}=\alpha_{e\tau} = 0$. Then, while fitting the ``observed'' event spectrum 
 with the opposite hierarchy, we introduce $\alpha_{e\mu}$ or $\alpha_{e\tau}$ (one at-a-time) in the 
 fit and marginalize over it in the range of $10^{-55}$ to $10^{-52}$ along with other oscillation 
 parameters. During this analysis, we find that the mass hierarchy sensitivity of ICAL gets reduced 
 very marginally by around 5\%.

\section{Summary and Conclusions}
\label{sec:conclusion}

The main goal of the proposed ICAL experiment at INO is to measure the neutrino mass 
hierarchy by observing the atmospheric neutrinos and antineutrinos separately and 
making use of the Earth matter effects on their oscillations. Apart from this, ICAL 
detector can play an important role to unravel various new physics scenarios beyond 
the SM (see Refs.\,\cite{Dash:2014sza,Dash:2014fba,Chatterjee:2014oda,
Chatterjee:2014gxa,Choubey:2015xha,Behera:2016kwr,Choubey:2017eyg,Choubey:2017vpr}). 
In this paper, we have studied in detail the capabilities of 
ICAL to constrain the flavor-dependent long-range leptonic forces mediated by 
the extremely light and neutral bosons associated with gauged $L_e-L_\mu$ or $L_e-L_\tau$ 
symmetries.  It constitutes a minimal extension of the SM preserving its renormalizibility 
and may alter the expected event spectrum in ICAL. As an example, the electrons inside 
the sun can generate a flavor-dependent long-range potential $V_{e\mu/e\tau}$  at the 
Earth surface, which may affect the running of oscillation parameters in presence of 
the Earth matter. Important point to note here is that for atmospheric neutrinos, 
$\Delta m^2/2E\sim$ 2.5$\times$ $10^{-13}$ (assuming $\Delta m^2 \sim 2.5 \times 
10^{-3}$ eV$^2$ and $E$ = 5 GeV), which is comparable to 
$V_{e\mu/e\tau}$ even for $\alpha_{e\mu/e\tau}\sim 10^{-52}$, and can influence the atmospheric 
neutrino experiments significantly. Also, for a wide range of baselines accessible in 
atmospheric neutrino experiments, the Earth matter potentials ($V_{CC}$)  
are around $10^{-13}$ eV (see table\,\ref{tab:comp-values}), suggesting that $V_{CC}$ can interfere with $V_{e\mu/e\tau}$ 
and $\Delta m^2_{31}/2E$, and can modify the oscillation probability substantially. 
In this article, we have explored these interesting possibilities in the context of the 
ICAL detector. 

After deriving approximate analytical expressions for the effective 
neutrino oscillation parameters in presence of $V_{CC}$ and $V_{e\mu/e\tau}$, 
we compare the oscillation probabilities obtained using our analytical expressions 
with those calculated numerically. Then, we have studied the impact of 
long-range forces by drawing the neutrino oscillograms in $E_\nu$ and $\cos\theta_\nu$ plane 
using the full three-flavor probability expressions with the varying Earth matter densities 
based on the PREM profile\,\cite{PREM:1981}. We have also presented the expected 
event spectra and total event rates in ICAL 
with and without long-range forces. As non-zero $\alpha_{e\mu}$ and $\alpha_{e\tau}$ can 
change the standard 3$\nu$ oscillation picture of ICAL significantly, we can expect to 
place strong limits on these parameters if ICAL do not observe a signal of LRF in 
oscillations. The expected upper bound  on $\alpha_{e\mu/e\tau}$ from 
ICAL is $<1.2\times 10^{-53}$ ($1.75\times 10^{-53}$ ) at 90$\%$ (3$\sigma$) C.L. with 
500 kt$\cdot$yr exposure and NH as true hierarchy. ICAL's limit at 90$\%$ C.L. 
on $\alpha_{e\mu}$ ($\alpha_{e\tau}$) is $\sim 46$ (53) times better than the existing limit  
from the Super-Kamiokande experiment. Here, we would like to mention that if the range 
of LRF is equal or larger than our distance from the Galactic Center, then the collective 
long-range potential due to all the electrons inside the Galaxy needs to be taken into
account\cite{PhysRevD.75.093005}. In such cases, ICAL can be sensitive to even lower values 
of $\alpha_{e\mu/e\tau}$. We hope that our present work can be an 
important addition to the series of interesting physics studies which can be performed
using the proposed ICAL detector at the India-based Neutrino Observatory. 

\section{Acknowledgment}

This work is a part of the ongoing effort of INO-ICAL Collaboration to study various physics 
potentials of the proposed ICAL detector. Many members of the Collaboration have contributed
for the completion of this work. We would like to thank A. Dighe, A.M. Srivastava, P. Agrawal,
S. Goswami, P.K. Behera, D. Indumathi for their useful comments on our work. 
We acknowledge S.S. Chatterjee for many useful discussions during this work. 
We thank A. Kumar and B.S. Acharya for helping us during the INO Internal Review Process. 
A.K. would like to thank the Department of Atomic Energy (DAE), Government of India 
for financial support. S.K.A. acknowledges the support from 
DST/INSPIRE Research Grant No. IFA-PH-12. 

\appendix

\section{Oscillation Probability with $L_e-L_\mu$ Symmetry}
\label{app-1}

\begin{figure}[htb!]
 \subfigure[]{\includegraphics[width=7.5cm]{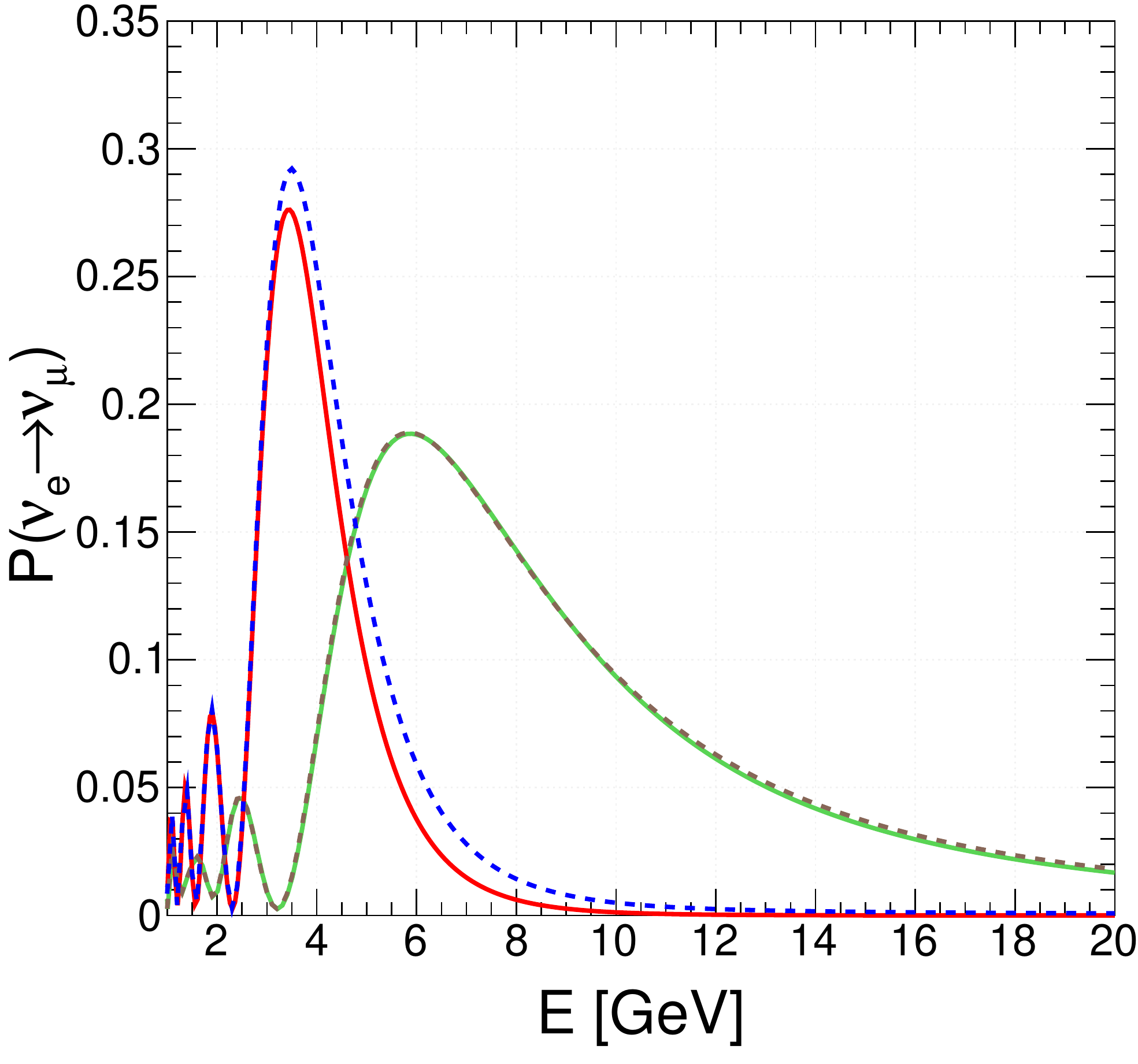}}
 \subfigure[]{\includegraphics[width=7.5cm]{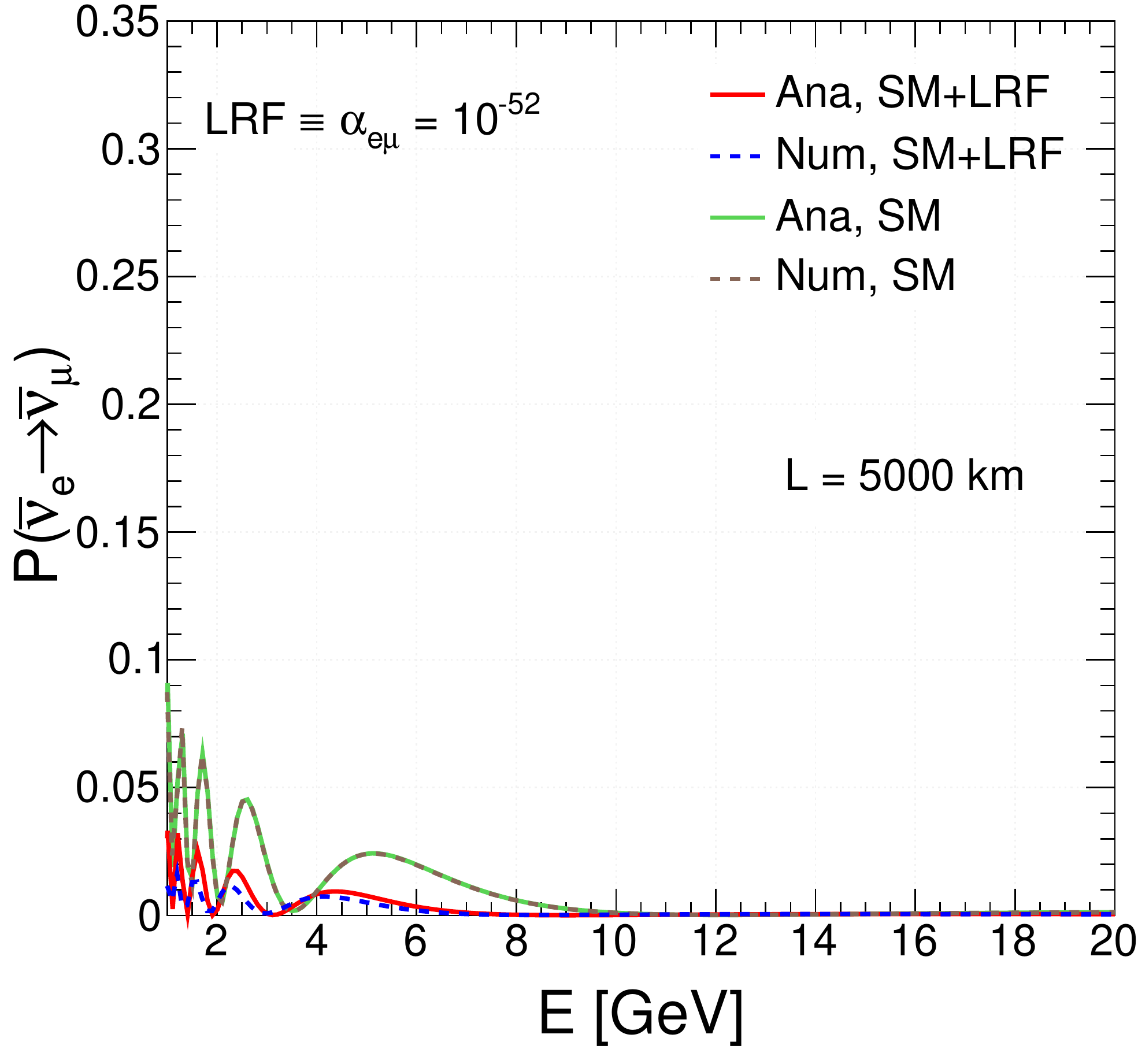}}
 \subfigure[]{\includegraphics[width=7.5cm]{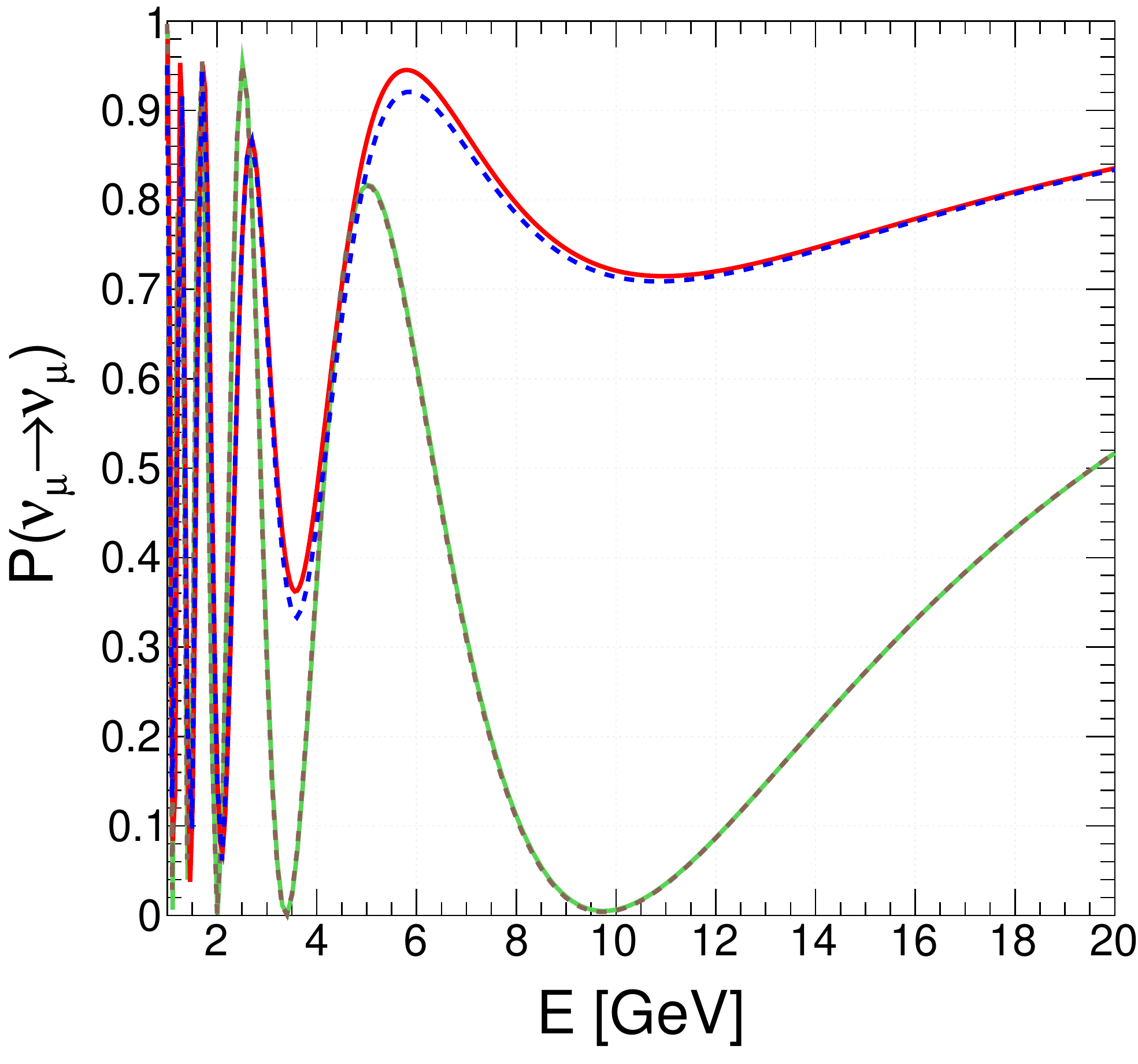}}
 \subfigure[]{\includegraphics[width=7.5cm]{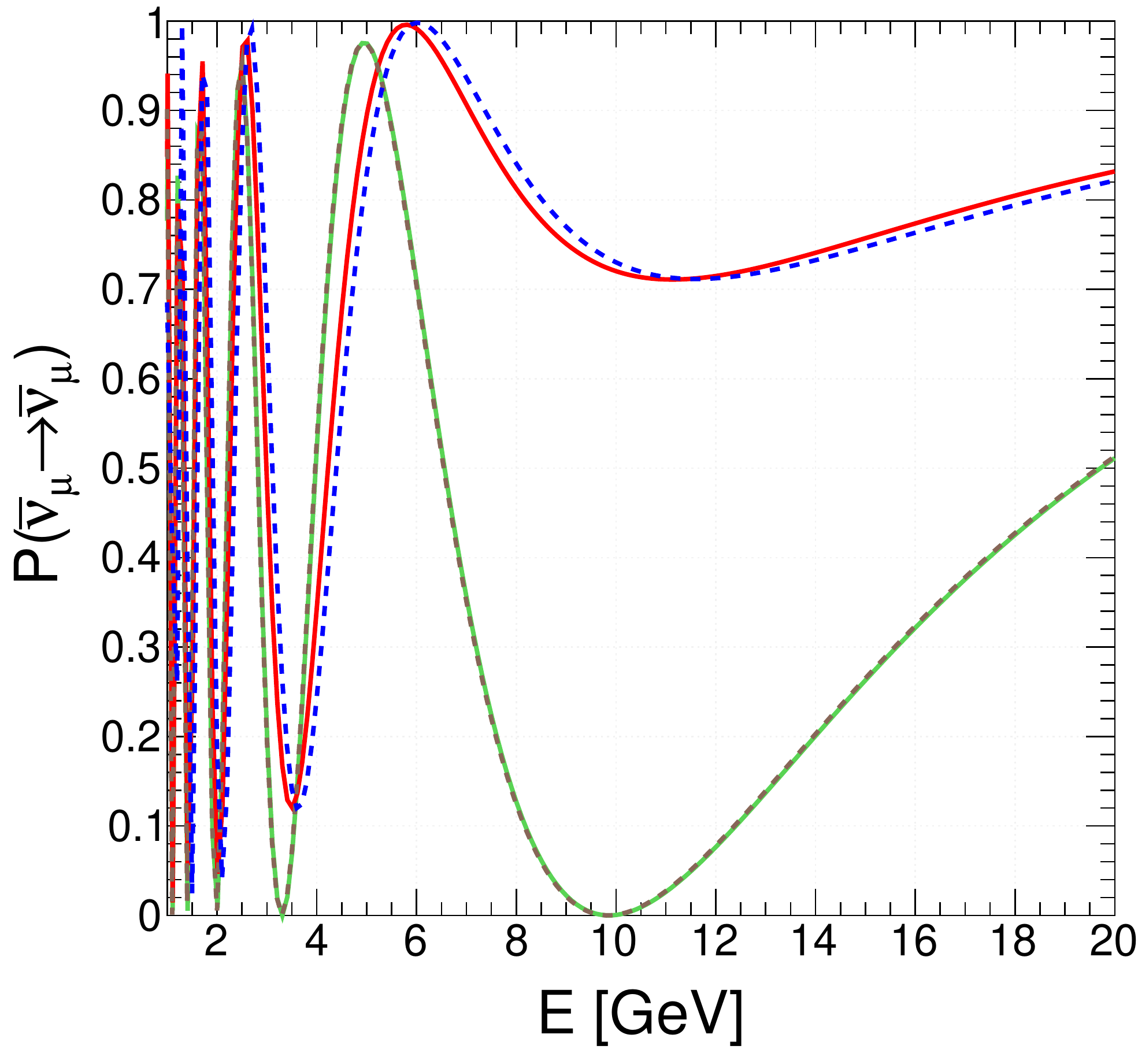}}
\mycaption{$\nu_e\rightarrow\nu_\mu$ ($\bar\nu_e\rightarrow\bar\nu_\mu$) transition probability for 
5000 km in upper left (right) panel assuming NH. In bottom left (right) panel, we show $\nu_\mu\rightarrow
\nu_\mu$ ($\bar\nu_\mu\rightarrow\bar\nu_\mu$) survival probability. In all the panels, we 
compare our analytical expressions (solid curves) to the exact numerical results (dashed curves) 
for the SM and SM\,+\,LRF cases. For LRF, we consider $\alpha_{e\mu}=10^{-52}$. }
\label{fig:emu-prob-nu-anu-analyt}
\end{figure}
Fig.\,\ref{fig:emu-prob-nu-anu-analyt} shows approximate $\nu_e \rightarrow\nu_\mu$ 
($\bar\nu_e\rightarrow\bar\nu_\mu$) oscillation probabilities in the top left (right) 
panel as a function of $E$ against the exact numerical results considering 
$L = 5000$ km and NH. We repeat the same for $\nu_\mu \rightarrow\nu_\mu$ ($\bar\nu_\mu \rightarrow \bar\nu_\mu$) 
survival channels  in bottom left (right) panel. We perform these comparisons among 
analytical (solid curves) and numerical (dashed curves) cases for both the SM and SM\,+\,LRF 
scenarios considering our benchmark choice of $\alpha_{e\mu}=10^{-52}$.
For the SM case ($\alpha_{e\mu}=0$), the approximate results match exactly with numerically obtained 
probabilities. Analytical expressions also work quite well in presence of $L_e-L_\mu$ symmetry, 
and can produce almost accurate $L/E$ oscillation patterns.

\bibliographystyle{JHEP}
\bibliography{nsi-lrf-references.bib}

\end{document}